\documentclass[structabstract,english]{aa}                                    

\usepackage[T1]{fontenc}
\usepackage[latin9]{inputenc}
\usepackage{array}
\usepackage{amsmath}
\usepackage{graphicx}
\usepackage{amssymb}
\usepackage{esint}
\usepackage[authoryear]{natbib}

\providecommand{\tabularnewline}{\\}

\usepackage{babel}

\begin{document}

\title{Low velocity shocks: signatures of turbulent dissipation in diffuse irradiated gas.}

\subtitle{}

\author{P. Lesaffre\inst{1} 
\and G. Pineau des Forêts\inst{2,1} 
\and B. Godard \inst{3} 
\and P. Guillard \inst{4} 
\and F. Boulanger\inst{2} 
\and E. Falgarone\inst{1} }
  
\institute{ENS, LERMA, UMR 8112, CNRS, Observatoire de Paris, 24 rue Lhomond 75005 Paris, France LRA/ENS 
\and Institut d'Astrophysique Spatiale (IAS), UMR 8617, CNRS, Université Paris-Sud 11, Bâtiment 121, 91405 Orsay Cedex, France 
\and Departamento de Astrof\'isica, Centro de Astrobiolog\'ia, CSIC-INTA, Torrej\'on de Ardoz, Madrid, Spain
\and Spitzer Science Center (SSC), California Institute of Technology, MC 220-6, Pasadena, CA 91125, USA }

 \date{Received ...; accepted ...}

\abstract
{ Large scale motions in galaxies (supernovae explosions, galaxy
collisions, galactic shear ...) generate turbulence, allowing a fraction of the 
available  kinetic energy to cascade 
down to small scales before being dissipated.  }
{We establish and quantify the diagnostics of
turbulence dissipation in mildly irradiated diffuse gas in the specific
context of shock structures. }
{We incorporate the basic physics of photon-dominated
regions into a state-of-the-art steady-state shock code. We examine
the chemical and emission properties of mildly irradiated (G$_{0}$=1)
magnetised shocks in diffuse media ($n_{\rm H}=10^{2}\mbox{ to }10^{4}\,\mbox{cm}^{-3}$
) at low to moderate velocities (from 3 to 40 km.s$^{-1}$). }
{ The formation of some molecules relies on endoergic reactions. 
In J-shocks, their abundances are enhanced by several orders of magnitude for 
shock velocities as low as 7 km.s$^{-1}$.  Otherwise most
chemical properties of J-type shocks vary over less than an order of
magnitude between velocities from about 7 to about 30 km.s$^{-1}$,
where H$_{2}$ dissociation sets in.  C-type shocks display a more
gradual molecular enhancement as the shock velocity increases. 

We quantify the energy flux budget (fluxes of kinetic, radiated and
magnetic energies) with emphasis on the main cooling lines of the cold interstellar medium.
Their sensitivity to shock velocity is such that it allows observations to constrain
statistical distributions of shock velocities.

 We fit various probability distribution functions (PDFs) of shock velocities
to spectroscopic observations of the galaxy-wide shock in
Stephan's Quintet (SQ) and of a Galactic line of sight sampling
diffuse molecular gas in Chamaeleon.  In both cases, low velocities
bear the greatest statistical weight and the PDF is consistent with a bimodal
distribution. In the very low velocity shocks (below 5 km.s$^{-1}$),
dissipation is due to ion-neutral friction which powers H$_2$  low energy transitions and
atomic lines. In moderate velocity shocks (20 km.s$^{-1}$ and above),
the dissipation is due to viscous heating and accounts for most of the
molecular emission. In our interpretation a significant fraction of
the gas on the line of sight is shocked (from 4\% to 66\%). For example, C$^{+}$ emission may trace shocks in UV irradiated gas where C$^+$ is the dominant carbon species.}
 { Low and moderate velocity shocks are important in shaping the chemical composition and
 excitation state of the interstellar gas. This allows to probe the statistical distribution
of shock velocities in interstellar turbulence.
}


\authorrunning{Lesaffre et al.}
\titlerunning{Low-velocity shocks in diffuse irradiated gas}

\maketitle

\section{Introduction}

{\bf Bulk motions in galaxies generated by e.g. supernovae explosions,
galaxy collisions and galactic shear drive turbulence in the cold
interstellar medium (ISM). A fraction of the available kinetic energy
cascades down to smaller scales and lower velocities. This is
spectacularly illustrated by the observations of the galaxy collision
in Stephan's Quintet (SQ). The relative velocities of the galaxies (on
the order of 1000 km.s$^{-1}$) would be expected to dissipate in high
velocity shocks, thus creating a warm and hot plasma devoid of
molecules. However, one observes that H$_2$ cooling is greater than the X-ray luminosity \citep{2006ApJ...639L..51A, 2010ApJ...710..248C}. This demonstrates
that energy dissipation in the interstellar space involves an energy
cascade and molecular gas cooling. Moreover, the H$_2$ excitation
diagram in SQ implies a distribution of tempatures much above the
equilibrium temperature set by UV and cosmic ray heating. The same
holds on much smaller scales in the case of the diffuse ISM in the
Solar Neighbourhood \citep{2002A&A...391..675G,2011ApJ...743..174I}.
The range of gas temperatures can only be accounted for if the dissipation heats
the gas through spatially localised events
\citep{2005A&A...433..997F}. This idea is also supported by
observations of chemical species in the diffuse ISM such as CH$^+$
and SH$^+$ which cannot be reproduced by UV alone \citep{2008A&A...483..485N,2012A&A...540A..87G}.

These dissipation processes have yet to be studied. Dissipative
structures could take the form of e.g. shocks, vortices, current
sheets, shear layers and most likely involve the magnetic field.
This broad variety of processes, their different time scales and
therefore different impacts onto the global gas energetics and
chemistry, call for simplified models to quantify observational
diagnostics.  Spectroscopy of
molecular H$_{2}$ with Copernicus in the diffuse ISM motivated the
first MHD shock models propagating in diffuse gas. These first models
were also key to provide a first interpretation of CH$^{+}$ in the
diffuse ISM
\citep{1971MNRAS.153..145M,1983ApJ...264..485D,1986MNRAS.218..729F,2002A&A...389..993G}. ISO-SWS
observations \citep{2005A&A...433..997F} of the pure rotational lines
of H$_2$ in diffuse gas were also interpreted in the framework of mild
turbulent dissipation (low velocity MHD shocks and/or velocity shears
in small scale vortices). More recently, \citet{2009A&A...495..847G}
have modeled turbulent dissipation bursts in the diffuse ISM as
 vortices at very small scales and computed
the chemical signatures they imprint on the gas.

In the present work, we use an updated version of the magnetised shock
models of \citet{2003MNRAS.343..390F} to quantify composition and
cooling lines for a range of shock velocities (from 3 to 40
km.s$^{-1}$) including very low velocities which were not thoroughly
considered before. We compute grids of such shocks for two strengths
of magnetic field and for three different densities. Our preshock
conditions are representative of the cold diffuse ISM. Hence, we have
incorporated the treatment of mild irradiation which includes UV
heating but also impacts the chemistry through photo-ionisation and
photo-dissociation.

The shock velocity can be a crucial parameter even at very low
velocity.  As it happens, the generation of molecules in the
interstellar medium (ISM) is initiated by the formation of the H$_{2}$
molecule on dust grains surfaces.  Then, the formation of more
elaborate molecules relies on two main paths.  First, one can add a proton to
H$_2$ to form H$_3^+$ and then transfer the extra proton to a single
atom (C, O, S and Si, with the notable exception of N). But
H$_3^+$ is usually much less abundant than H$_2$. Second, one can directly
exchange an atom with one proton of the H$_2$ molecule. But these
reactions are subject to energy barriers with high characteristic
temperatures (such as 2980K in the case of the oxygen
atom). Hence both paths are difficult and the observed molecular complexity
of the ISM stands as a puzzle. Nevertheless, temperatures required 
to open the second path are obtained in low
velocity shocks (7.5 km.s$^{-1}$ is enough to reach 3000K). These
shocks may therefore play an important role in shaping the molecular
chemistry of the ISM.

The results of the present models may contribute to the interpretation
of the observations of the full set of cooling lines of the cold
neutral medium (CII, CI, OI, H$_{2}$, CO and H$_{2}$O) in galaxies
observed with the {\it Spitzer} and {\it Herschel} space telescopes.  To
illustrate this, we build models from statistical distributions of
shock velocities which we compare to observed data in the SQ and in
a line of sight in the Chamaeleon \citep{2002A&A...391..675G}.  We use our best fit models to discuss the impact of
kinetic energy dissipation on the physical state and chemistry of
the molecular gas in these two sources.  We also make predictions on other
diagnostic lines to be observed with {\it Herschel}.
} 
We present the numerical method for our shock models in section 2,
with the properties of our grid of shocks in section 3. In section 4,
we use our results to interpret the H$_{2}$-excitation diagram
observed in SQ and Chamaeleon. We summarize and discuss our results in
section 5.

\section{Numerical method}

The models we present in this paper are based on the plane-parallel steady-state
shock code implemented in \citet{2003MNRAS.343..390F}. We work from
their version and include few more refinements mainly in order to
deal with moderate irradiation.

\subsection{Radiation }

We start from the reactions network used in \citet{2003MNRAS.343..390F}
and include the relevant photo-reactions. The photo-reactions have
their rates of the form\begin{equation}
R=\alpha G_{0}e^{-\beta A_{v}}\end{equation}
where $\alpha$ and $\beta$ are constants. We use an incident field
equal to the standard interstellar radiation field (ISRF, \citealt{1978ApJS...36..595D}),
thus taking $G_{0}=1$. The extinction $A_{v}$ is integrated along
the model from the pre-shock where we use a value of $A_{v}=A_{v0}=0.1$ which 
models extinction from the irradation source by a ``buffer'' of matter. 
The value of $A_{v0}$ is a parameter of the problem which selects how much mass
is contained in this buffer. 
The local value for the extinction is then computed as \begin{equation}
A_{v}=A_{v0}+\intop_{x_{0}}^{x}\mathrm{d}z\,\sigma_{\mathrm{g}}n_{\rm H}\end{equation}
where  {\bf $\sigma_{\mathrm{g}}=5.34\,\times 10^{-22}\,\mathrm{cm^{3}}\mathrm{pc}^{-1}$ is the effective extinction per H nuclei column-density, $n_{\rm H}$ is the local density of H nuclei,  }
$x$ is the current position and $x_{0}$ is the position at
the entrance of the preshock (ie: the point where we start our simulations). Our shocks are hence assumed to be irradiated
{}``backward'' compared to their direction of propagation. However
this matters only a little because the total extinction through these
shocks is small (on the order of $\Delta A_{v}=0.01\,(n_{\rm H}/100$
cm$^{-3}$), where $n_{\rm H}$ is the pre-shock density).  {\bf This setup is in fact exactly similar to the one used by \citet{2004ApJ...612..921B}. However they focus on the properties of a uniform pressure photon-dominated region (PDR) following the shock, whereas we stop our computation right behind the shock, and they consider pre-shock densities of $n_{\rm H}=1~$cm$^{-3}$ much lower than ours.}

The reaction rates for the photo-dissociation of H$_{2}$ and CO include
an additional factor $f_{\mathrm{shield}}$ to account for the shielding
and self-shielding. We use the tables of \citet{1996A&A...311..690L}
for CO and \citet{1996ApJ...468..269D} for H$_{2}$ with a Doppler
parameter  
\begin{equation}
b_{D}=\sqrt{u_{\rm th}^2 + u_{\rm turb}^2}
\end{equation}
 where $u_{\rm th}$ is the thermal velocity of H$_2$ molecules and
$u_{\rm turb}=1$ km.s$^{-1}$ accounts for microturbulence.  However,
we neglect the effect of velocity gradients on the self-shielding.
For instance:\begin{equation} R_{{\rm H}_{2}+h\nu}=(2.54\,\times
10^{-11}s^{-1})\, f_{\mathrm{shield}}e^{-6.3\, A_{v}}\end{equation}

where\begin{equation}
f_{\mathrm{shield}}=\frac{0.965}{(1+x/b_D)^{2}}+\frac{0.035\, e^{-8.5\,\times 10^{-4}\sqrt{1+x}}}{\sqrt{1+x}}\end{equation}
with $x=N({\rm H}_{2})/(5\,\times 10^{14}\mathrm{cm^{-2}})$. Here, the quantity
$N($H$_{2})$ is the column-density of H$_{2}$ molecules, computed
as for the extinction:\begin{equation}
N({\rm H}_{2})=N_{0}({\rm H}_{2})+\int_{x_{0}}^{x}\mathrm{d}z\, n({\rm H}_{2})\end{equation}
where $n({\rm H}_{2})$ is the local density of ${\rm H}_{2}$ molecules and $N_{0}({\rm H}_{2})$
is the quantity of  ${\rm H}_{2}$ molecules in the ``buffer'' which shields the medium from the external
radiation field. 
We match $N_{0}({\rm H}_{2})\simeq10^{20}\mathrm{cm}^{-2}~$ to
$A_{v0}=0.1$ by assuming this buffer is mainly molecular. Similarly we
define $N_0$(CO) for the CO self-shielding but we use $N_0$(CO)=0 as
$A_{v0}=0.1$ is usually well below the extinction at which CO is
self-shielded in standard models of photon-dominated regions (PDR).

\label{Irradiation}With this irradiation model, we integrated the
chemical and thermal evolution of a fluid parcel with fixed density
moving away from the radiation source at a constant velocity
$v_{\mathrm{PDR}}$ where the subscript PDR refers to photon-dominated
region.  The resulting spatial profile (the trace of the thermal and
chemical composition history of this fluid parcel along its path)
corresponds to a steady PDR front moving towards the irradiation
source at speed $v_{\mathrm{PDR}}$. For very small velocities we
recover a steady PDR front which compares quite satisfactorily to the
Meudon PDR code \citep{2006ApJS..164..506L} for $G_0=1$. {\bf In the frame of this comparison} we switched off grains and PAHs
 reactions {\bf in order to remain} closer to the Meudon PDR code setup. We note the PDR structure
is insensitive to the choice of \citet{1996A&A...311..690L} or
\citet{1996ApJ...468..269D} for H$_{2}$ photo-dissociation. This is
due to the fact that the rates differ only in regions where the
absolute value of the rates are small, hence unimportant. For the same
reason, we expect our results should poorly depend on the exact value
of the above mentioned Doppler parameter
$b_{D}$. \citet{1987ApJ...322..412B} also shows that this parameter is
not crucial in PDRs.
 
Compared to \citet{2003MNRAS.343..390F}, we refreshed the collision
rates of OI by H atoms with the rates computed by \citet{2007ApJ...654.1171A}.
These rates enter the computation of the atomic cooling due to O atoms.
Atomic and molecular cooling are otherwise identical to those employed in \citet{2003MNRAS.343..390F}. For instance, we use the molecular cooling rates tabulated in \citet{1993ApJ...418..263N}
for the cooling by CO and H$_{2}$O.

\subsection{${\rm H}_{2}$-excitation}

We follow the time-dependent excitation of ${\rm H}_{2}$ along the shock
structure as in \citet{2003MNRAS.343..390F}. Here, we include the
treatment of the ${\rm H}_{2}$ population for the 149 lowest energy levels.
We check that this allows to compute the ${\rm H}_{2}$ cooling accurately
for shock velocities up to at least 40 km.s$^{-1}$ for the range of densities
we consider in this work (see also \citealp{2003MNRAS.343..390F}). 

We also use the knowledge of the population of  the 8 lowest H$_{2}$
rotational levels to compute more accurately the rate of the reaction
C$^{+}$ + H$_{2}$. To compute the rate state by state, we use \citet{1987JChPh..87..350G}
(as in \citealt{2010ApJ...713..662A}, line 2 of table 1) for H$_{2}$
levels with $J=0...7$ and $v=0$, and we use \citet{1997JChPh.10610145H}
(as in \citet{2010ApJ...713..662A}, line 3 of table 1) for the other levels.

\subsection{Grains}

As in \citet{2003MNRAS.343..390F}, we treat adsorption, 
collisional sputtering and collisional desorption of molecules from and onto grains.
However, unlike them, we take pre-shock conditions without ice mantles.

As in \citet{2003MNRAS.343..390F} we account for the charge of grains by including all charge exchanges involving grains and electron detachment by {\bf cosmic ray induced} secondary photons. The
heating due to the photo-electric effect is included, however we discard the detachment of electrons due to the radiation field in the chemical network.

\section{Grid of Models}
\label{models}

\subsection{Numerical protocole}

For each value of our parameters described in the next section, we integrate the steady-state equations
of multi-fluid MHD shocks (\citealt{1985MNRAS.216..775F,1990CoPhC..58..169H})
from entrance conditions at thermal and chemical equilibrium (see
details of the pre-shock chemical conditions below). As stated above,
we account for the changes in the properties of the irradiation field
due to the increasing absorption as we penetrate deeper into the shock
structure. The computation is stopped when the temperature decreases
back to 20\% above the temperature of the pre-shock.

\subsection{Choice of parameters}

We tune most of our parameters to the typical conditions encountered
in the dilute interstellar gas in our galaxy. We list the main physical
parameters of our model in table \ref{Flo:parameters-1}. 

\begin{table}
\label{Flo:parameters-1}\begin{tabular}{ccc}
Parameter & Value & Comment \tabularnewline
\hline
\hline 
$n_{\rm H}$ & $10^{2},$ $10^{3}$ or $10^{4}$cm$^{-3}$ & H nuclei pre-shock density\tabularnewline
$u$ & $3$ to $40$ km.s$^{-1}$ & Shock velocity\tabularnewline
$b=B/n{}_{H}^{1/2}$ & 0.1 or 1 & {\bf Dimensionless} magnetic field\tabularnewline
$A_{v0}$ & 0.1 & Buffer extinction\tabularnewline
$N_{0}({\rm H}_{2})$ & $10^{20}$cm$^{-2}$ & Buffer ${\rm H}_{2}$ column-density\tabularnewline
$N_{0}({\rm CO})$ & 0 & Buffer ${\rm CO}$ column-density\tabularnewline
$G_{0}$ & 1 & External radiation field\tabularnewline
$\zeta$ & $3\,\times 10^{-17}$s$^{-1}$ & Nominal cosmic rays flux\tabularnewline
OPR & 3 & Preshock H$_{2}$ ortho/para ratio\tabularnewline
\hline
\end{tabular}

\caption{Values for the main physical parameters in our models.}

\end{table}

We specify indirectly the strength of the magnetic field transverse to the shock speed by using
the {\bf non-dimensional} parameter $b=(B/1\mu\mathrm{G})/\sqrt{n_{\rm H}/\mathrm{cm^{3}}}$
which \citet{2010ApJ...725..466C} observe to take values from $b=0.1$
to $b=1$ for our range of densities. We later refer to these values
by highly magnetized shocks for $b=1$ and weakly magnetized shocks for
$b=0.1$.

For both of these assumptions on the magnetic field, we build three
grids of models, one for each pre-shock density between $n_{\rm H}=10^{2}~\mathrm{cm^{-3}}$,
$n_{\rm H}=10^{3}~\mathrm{cm^{-3}}$ and $n_{\rm H}=10^{4}~\mathrm{cm^{-3}}$.
{\bf  In
the following, we will use the term \emph{low} velocity for shocks
below 20 km.s$^{-1}$, \emph{moderate} velocity above 20 km.s$^{-1}$ and
\emph{high} velocity at and above 40 km.s$^{-1}$.}
Each grid spans a range of velocities from 3 up to 40 km.s$^{-1}$ with a
step of 0.5 km.s$^{-1}$, for an integration time of about 6 hours per grid on a typical workstation.
We choose our minimum velocity of 3 km.s$^{-1}$ above the Alfvén speed
in the neutral gas in the $b=1$ case:\begin{equation}
v_{A}=\frac{B}{\sqrt{4\pi\rho}}=\frac{b}{\sqrt{4\pi(\mu_{H}/\mathrm{a.m.u.})}}=1.85\, b\mathrm{\, km}/\mathrm{s}\end{equation}
 where the mean mass per H nucleus is $\mu_{H}=\rho/n_{\rm H}=1.4$ a.m.u.
. We choose the upper bound of 40 km.s$^{-1}$ because of the limitations
of our shock models\footnote{Indeed, the code does not treat doubly ionised species and
radiative emission from the post-shock itself which might affect pre-shock
conditions at high densities (see \citealt{2010MNRAS.406.1745F}).} .

\begin{table}
\begin{tabular}{ccccc}
 & \multicolumn{4}{c}{Fractional value $n(X)/n_{\rm H}$}\tabularnewline
Element $X$ & total & gas & grain cores & PAHs\tabularnewline
\hline
\hline 
H & 1 & 1 &  & 1.80(-5)\tabularnewline
He & 0.1 & 0.1 &  & \tabularnewline
C & 3.56(-4) & 1.38(-4) & 1.63(-4) & 5.40(-5)\tabularnewline
O & 4.42 (-4) & 3.02 (-4) & 1.40 (-4) & \tabularnewline
N & 7.94(-5) & 7.94(-5) &  & \tabularnewline
Mg & 3.70(-5) &  & 3.70(-5) & \tabularnewline
Si & 3.70(-5) & 3.37(-6) & 3.37(-5) & \tabularnewline
S & 1.86(-5) & 1.86(-5) &  & \tabularnewline
Fe & 3.23(-5) & 1.50(-8) & 3.23(-5) & \tabularnewline
\hline
\end{tabular}
\caption{Elemental composition of the pre-shock gas (as in \citealp{2009A&A...495..847G}).
Initial conditions assume no ice mantles (bare cores). Note the gas
to dust mass ratio is $\rho_{\mbox{gas}}/\rho_{\mbox{dust}}\simeq180$. The numbers in the parentheses denote the powers of ten.\label{Flo:composition}}

\end{table}

At each fixed pre-shock density, we first evolve the gas chemically
and thermally during $10^{7}$ years, which brings the gas
close to thermal and chemical equilibrium. We start with the same
elemental abundances as in \citet{2009A&A...495..847G}, which are
summarised in table \ref{Flo:composition}. Note the gas phase abundance
in the pre-shock is set to a tenth of the Si locked in
the grains cores, {\bf consistent with observations \citep{2009ApJ...700.1299J}}. We also
include PAHs {\bf with} a fraction $n(\mbox{PAH})/n_{\rm H}=10^{-6}$
of the representative species $\mbox{C}_{54}\mbox{H}_{18}$. Without irradiation ($G_0=0$), PAHs 
influence the ionisation degree of the gas and the charge-ion coupling. 
However with mild irradiation (at $G_0=1$) it turns out that C$^+$ 
ions dominate the charge fluid and the role of PAHs is negligible. 
With our current irradiation parameters ($G_0$, extinction and H$_2$ buffer), the atomic
hydrogen fraction is $n({\rm H})/n_{\rm H}=7.9(-2),$ 1.3(-2) and
2.0(-3) in the preshock gas with respective densities $n_{\rm H}=10^{2}~$cm$^{-3}$,
$10^{3}~$cm$^{-3}$ and 10$^{4}~$cm$^{-3}$. We then use the resulting
state as our pre-shock conditions to run the shock model for each
velocity in turn.

\subsection{C- and J-type shocks}

Steady-state magnetized shocks in the interstellar medium are of one
of two kinds: J-type shocks in which the kinetic energy is dissipated
viscously in a very sharp velocity \emph{jump} (hence 'J') followed
by a thermal and chemical relaxation layer and C-type shocks in which
kinetic energy is \emph{continuously} (hence 'C') degraded {\bf into} heat and photons via ion-neutral
friction and cooling. {\bf \citet{1971MNRAS.153..145M} was first  to discover that high magnetic fields
can transfer kinetic energy to thermal energy in a continuous
manner and coined the term C-shock. \citet{1980ApJ...241.1021D}, \citet{1983ApJ...264..485D},
and \citet{1990ApJ...350..700R} then described the multifluid nature of
these shocks. C-shocks occur as long as the shock velocity remains below the propagation speed of the magnetic precursor. This} critical velocity above which a C-shock
cannot exist is the magnetosonic velocity \begin{equation}
v_{m}=\sqrt{c_{s}^{2}+v_{Ac}^{2}}\end{equation}
where $v_{Ac}=B/\sqrt{4\pi\rho_{c}}$ is the Alfvén speed in the \emph{charged}
fluid and $c_{s}$ is the speed of sound. The Alfvén speed in the charges depends both on the magnetic
field and on the inertia in the charged fluid.  Thanks to the irradiation field, 
the gas is well ionised and provides ample free electrons to stick onto the grains. As a result most
grains are charged {\bf negatively} in our models and hence they dominate the inertia
in the charged fluid. Moreover, \citet{2007A&A...476..263G} showed
that even the neutral grains spend an extensive fraction of their
time attached to the magnetic field and they should also be included
in the charged fluid inertia. We hence include all PAH and grains
in the computation of the magnetosonic speed. As a result, the Alfvén
speed in the charged fluid $v_{Ac}$ is larger than $v_{A}$ by a
factor $v_{Ac}\sim\sqrt{\rho/\rho_{d}}\, v_{A}$ where $\rho/\rho_{d}\sim180$
is the gas to dust mass ratio. The actual number in \emph{our} simulations
turns out to be $v_{m}=21.2$ km.s$^{-1}$, 21.7 km.s$^{-1}$ and 22.9 km.s$^{-1}$ for respectively
$n_{\rm H}=10^{2}$cm$^{-3}$, $10^{3}$cm$^{-3}$ and 10$^{4}$cm$^{-3}$
in the case $b=1$ and about ten times lower values for $b=0.1$. 

Shocks with speed greater than $v_{m}$ will always be J-shocks.
For instance, all shocks in our $b=0.1$ grid are J-shocks.
However, time-dependent shocks with velocities lower than $v_{m}$
could be C-shocks, J-shocks or even a combination of the two (see
\citealp{1998MNRAS.295..672C,2004A&A...427..147L,2004A&A...427..157L}).
Indeed, {\bf (i)} the transverse magnetic field could be lower for
various orientations of the field (see \citealp{1998MNRAS.298..507W}
for oblique C-shocks), {\bf (ii)} or the C-shock could be at an earlier development
of its structure where it still has a J-shock component (\citealt{1998MNRAS.295..672C},
\citealt{2004A&A...427..147L}), {\bf (iii)} or the final fate of the shock could
be a steady-state CJ-shock depending on the history of the shock (\citealt{2004A&A...427..147L},
\citealt{2004A&A...427..157L}). As a result, we build our grids of
models with C-shocks for shock velocities up to $v_{m}$ and we
use J-shocks for higher shock velocities. In the following, we
explore the properties of the resulting shocks.

\subsection{Weakly magnetized shocks (b=0.1)}
\label{modelsJ}
For high compression ratios (ie: Mach numbers much greater than 1), the maximum temperature
in a J-shock (obtained just behind the leading adiabatic shock front)
is 

\begin{equation}
T_{\mathrm{max}}=\frac{2(\gamma-1)}{(\gamma+1)^{2}}\frac{\mu}{k_{B}}u^{2}\label{eq:Tmax0}\end{equation}
where $\mu$ is the mean molecular weight of the gas, 
$k_B$ is the Boltzman constant and $u$ is {\bf the
shock velocity in the shock frame. Note the shock velocity is very close to the velocity difference between upstream and downstream
gas for high Alfvénic Mach number.} In particular, this relation shows that the peak temperature in a shock is proportional to the
square of the shock velocity.

\begin{figure}
(a)\includegraphics[width=8.8cm]{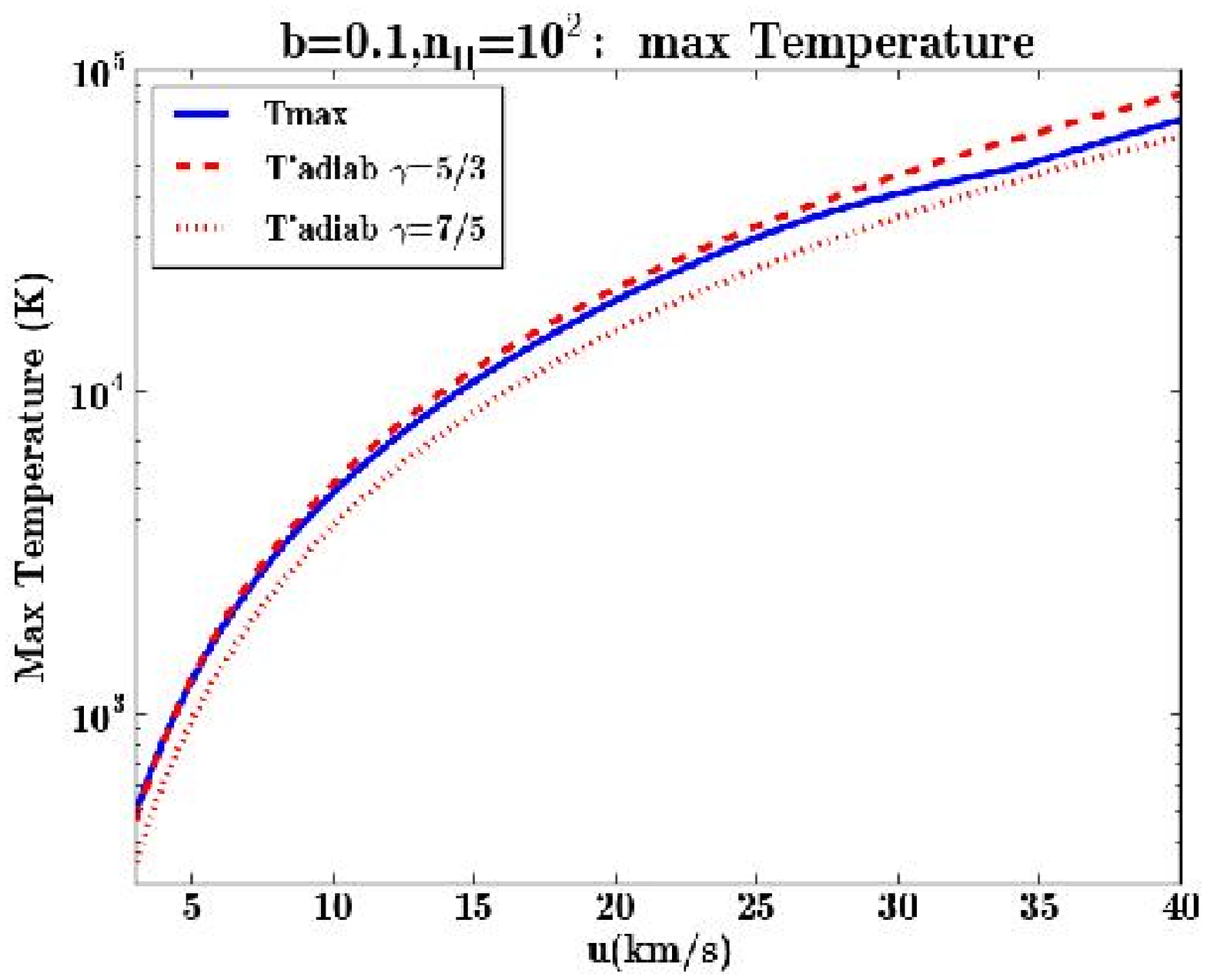}

(b)\includegraphics[width=8.8cm]{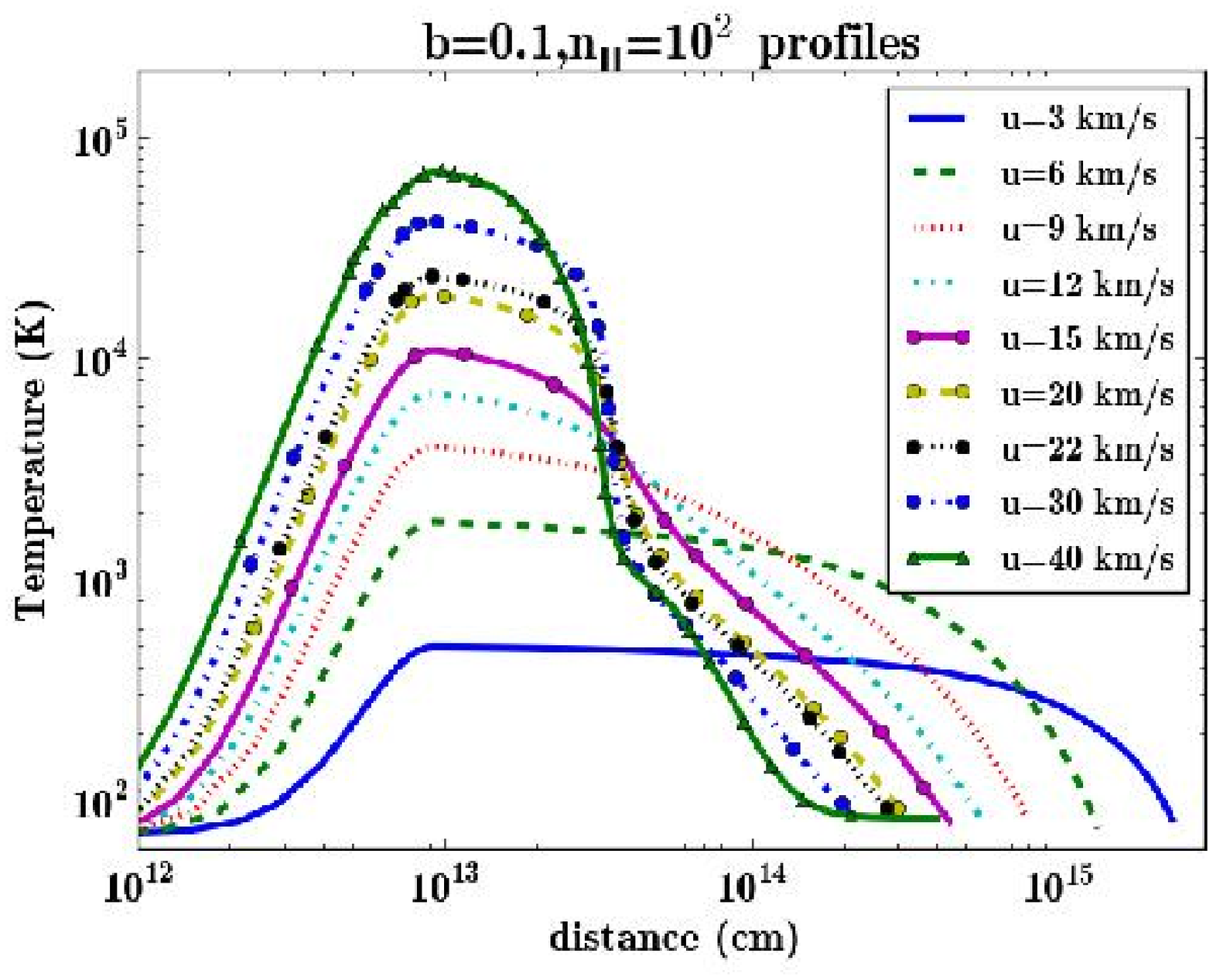}

\caption{(a) Maximum temperature in the \emph{weakly magnetized} shocks. (b)
Temperature profiles for some representative \emph{weakly magnetized}
shocks{\bf, the fluid flows from left to right with the pre-shock on the left and the post-shock on the right. Only J-shocks are present for this low value of the magnetic field}. $n_{\rm H}=10^2~$cm$^{-3}$. \label{Flo:b0.1.T}}

\end{figure}

{\bf For such dilute molecular gas, only the lowest energy levels of H$_2$ are populated and the appropriate value for $\gamma$ is $\gamma=5/3$. With}
$\mu=2.33$ a.m.u. (which corresponds to the above $\mu_{H}=1.4$ in our near fully molecular conditions)
 relation \eqref{eq:Tmax0} becomes:
\begin{equation}
T_{\mathrm{max}}=53\,\mathrm{K}\,(u/1\,\mathrm{km.s^{-1}})^{2}\label{eq:Tmax}\mbox{.}\end{equation}
 Figure \ref{Flo:b0.1.T}(a) displays
the above theoretical value and the actual maximum temperature reached
in our models for each velocity tested in our grid at $n_{\rm H}=10^{2}$ cm$^{-3}$. Both values remain
close to one another. At high temperature (high velocity), the maximum
temperature is closer to the value obtained in equation \eqref{eq:Tmax0}
with $\gamma=7/5$. This stems from the excitation of rotational levels of the H$_{2}$ molecules in the
adiabatic front. Note that we use a viscous length $\lambda=1/\sigma n_{\rm H}$
with a viscous cross-section $\sigma=3\,\times 10^{-15}\mbox{ cm}^{2}$ based
on H$_2$-H$_2$ collision data computed by \citet{1980JChPh..73.6153M} for a velocity dispersion
of 1 km.s$^{-1}$. We checked that the viscous front between the peak temperature
and the pre-shock is always adiabatic. {\bf Its width of 10$^{13}~$cm, visible on Figure \ref{Flo:b0.1.T}(b), corresponds to the viscous length for $n_{\rm H}=10^{2}$ cm$^{-3}$. This figure 
 surprisingly shows that higher velocity shocks return to pre-shock temperature
on a smaller distance than low velocity shocks.} However, the range of scales spans only
one order of magnitude from $2\,\times 10^{15}$ cm down to $2\,\times 10^{14}$
cm with all shocks with velocities higher than 20~km.s$^{-1}$ around the
scale of $2\,\times 10^{14}$ cm. Note that the range of flowing time through
these shock structures also spans about one order of magnitude, from
400 yr to 2000 yr. For higher densities, the picture is much the same and in particular
time scales vary little within one category (J-type or C-type) of shocks. However, shocks
which strongly dissociate H$_2$ (such as dense shocks at moderate velocity) have a wide
H$_2$ reformation plateau where H$_2$ reformation provides the necessary heat to keep
the gas warm \citep{2003MNRAS.341...70F}.

\begin{figure}
(a)\includegraphics[width=8.8cm]{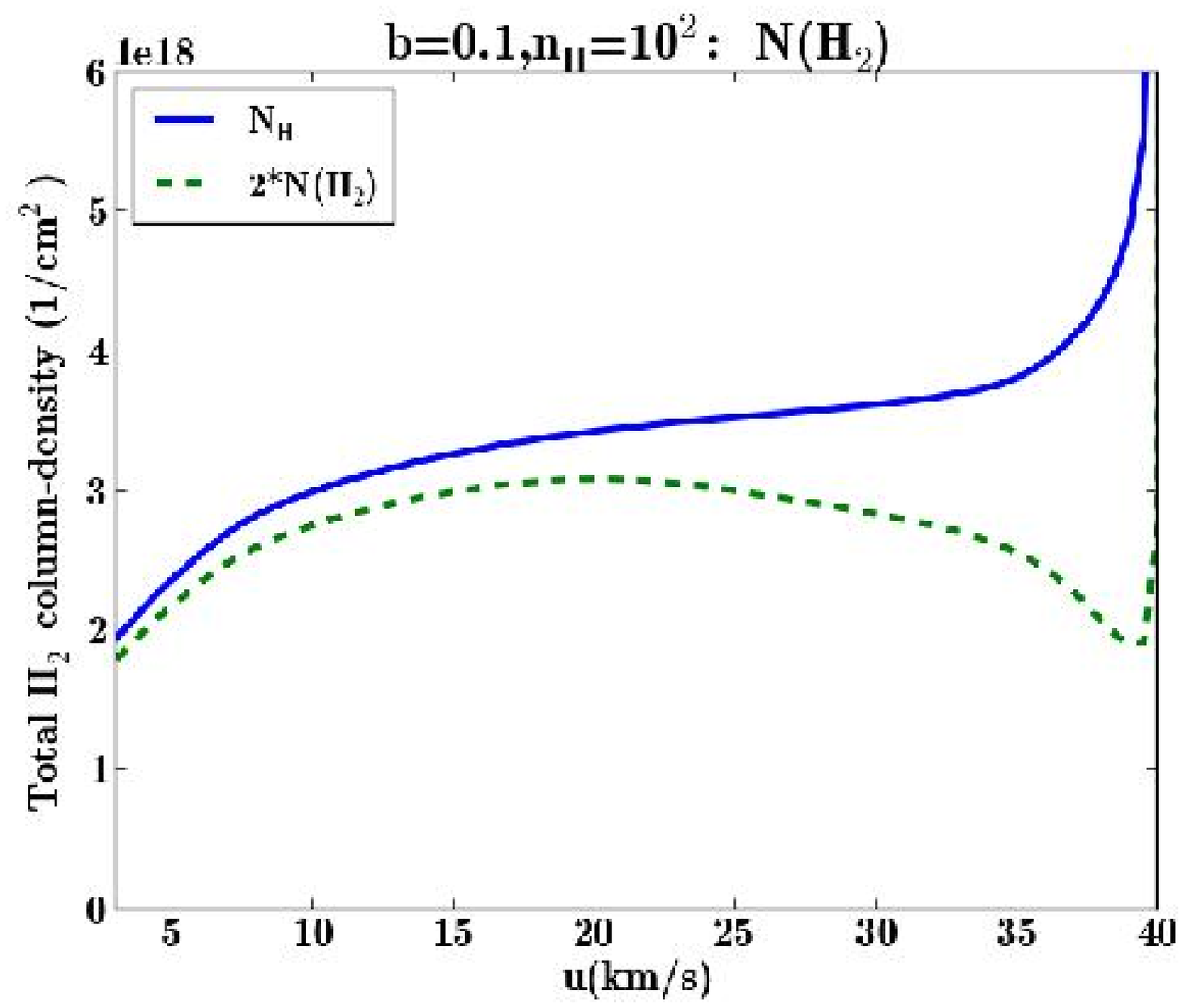}

(b)\includegraphics[width=8.8cm]{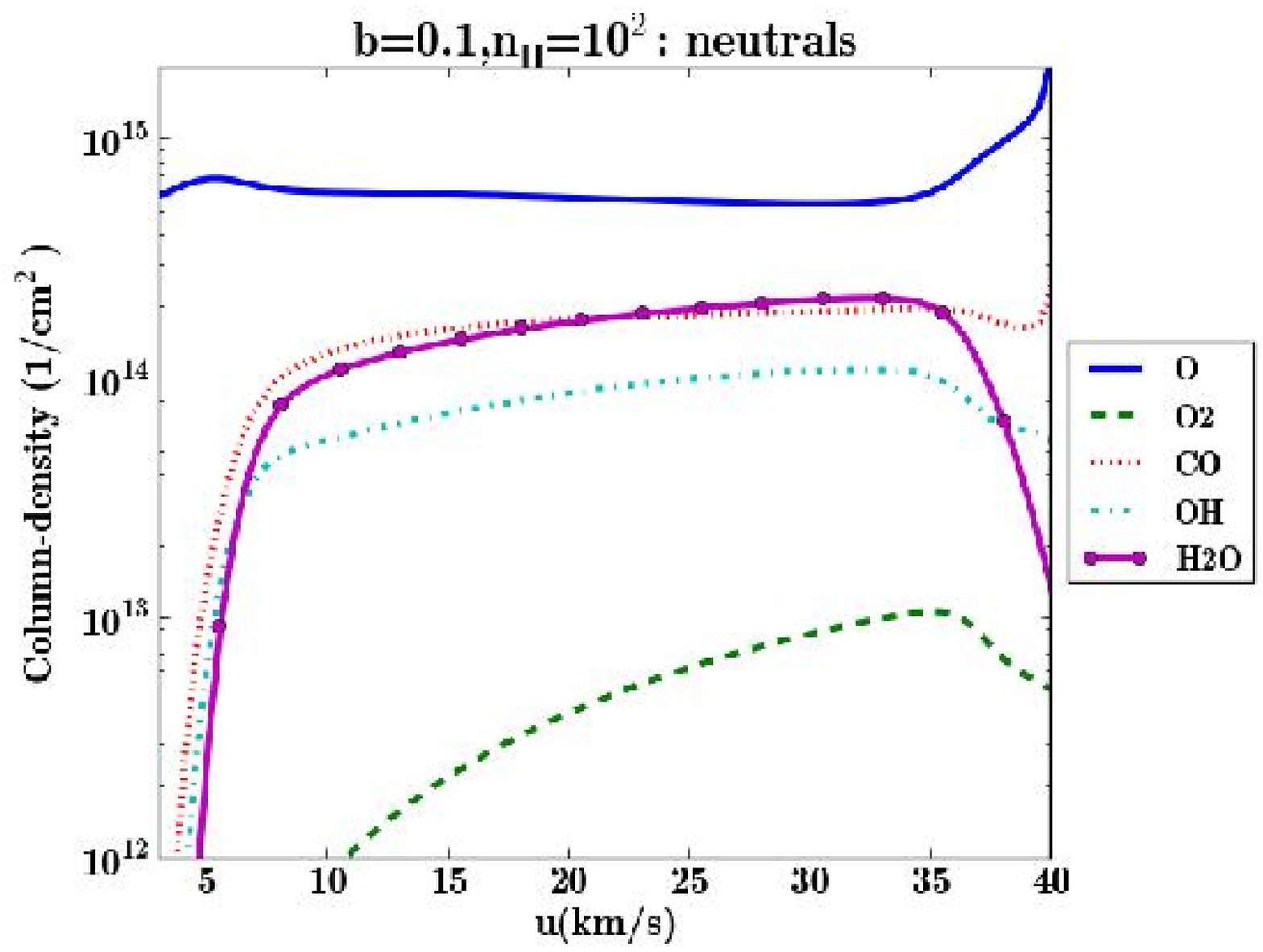}

(c)\includegraphics[width=8.8cm]{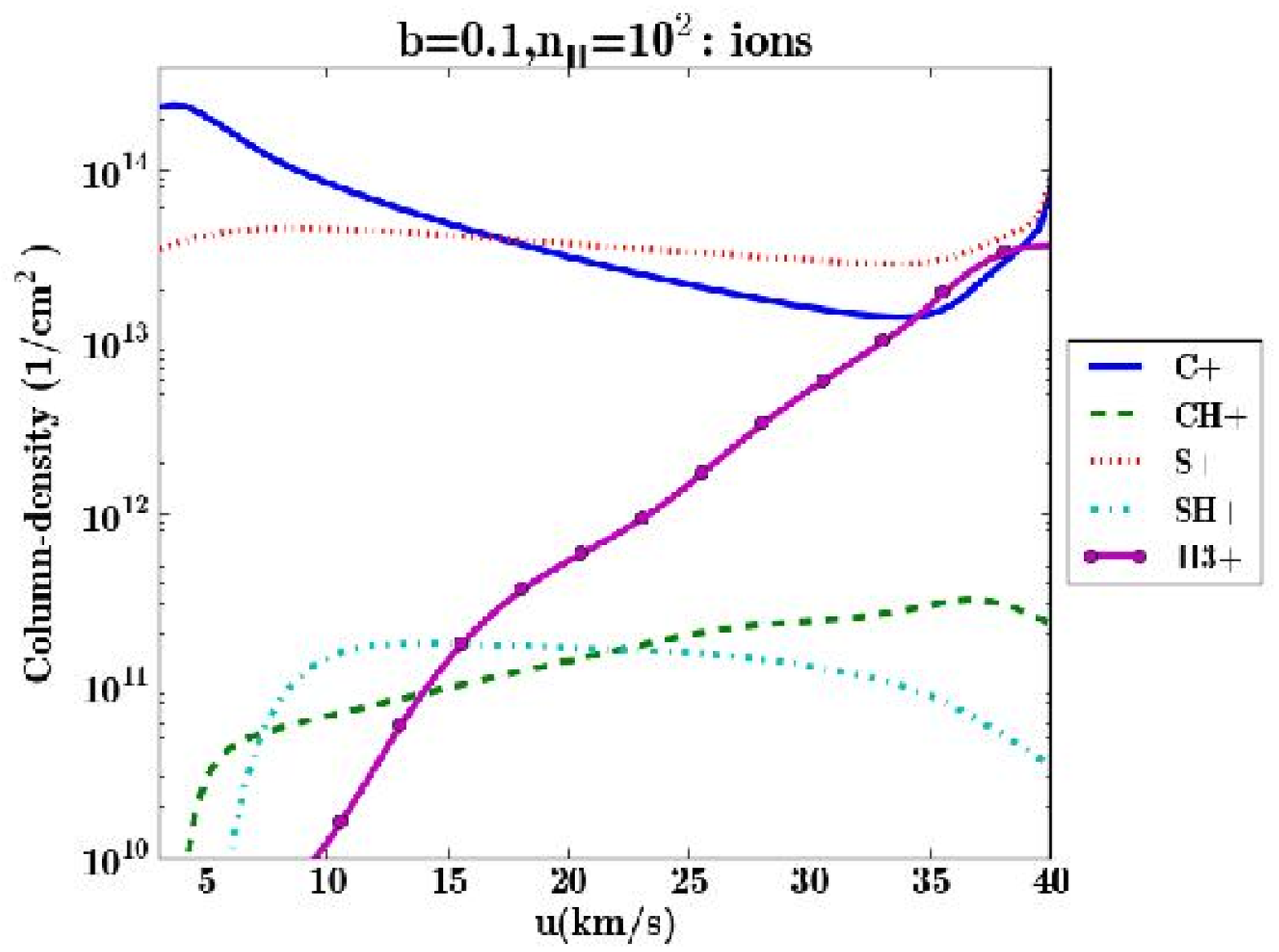}

\caption{For each shock velocity in our \emph{weakly magnetized }shocks ($b=0.1$), we
plot (a) the total H nuclei and H$_{2}$ molecules column-densities
(b) some neutrals of interest and (c) some ions of interest. $n_{\rm H}=10^2~$cm$^{-3}$. \label{Flo:b0.1}}

\end{figure}

Figure \ref{Flo:b0.1}(a) shows the total column-density of H nuclei
($N_{\rm H}$) across each shock in the grid. Also displayed is the 
total H$_{2}$ column-density. The total $N_{\rm H}$ column-density is surprisingly
constant around $N_{\rm H}$=3-4$\,\times 10^{18}$cm$^{-2}$ over the range
of velocities in our grid except at both ends. The increase of the compression factor at higher shock speed is compensated by the shorter cooling length.
 This result is in contrast
to the results obtained by \citet{1989ApJ...342..306H} in the velocity
range 30 to 150 km.s$^{-1}$ where the cooling column-density behind the shocks
is seen to vary greatly. Shocks with velocities below about 7 km.s$^{-1}$
have a lower than average total column-density, and shocks with velocities
above 35 km.s$^{-1}$ dissociate H$_{2}$ which decreases the amount of cooling
and increases the total column-density.

Figures \ref{Flo:b0.1}(b) and \ref{Flo:b0.1}(c) show the total
column-densities across the shock for various neutrals and ions.
Remember the end of each shock is defined when the temperature
decreases again under 20\% above the pre-shock temperature.  Since the
overall column-density is a slowly varying function of the shock
velocity (figure \ref{Flo:b0.1}(a)) these plots also give a good
estimate on how the average relative abundance of each of these
species varies with respect to the shock velocity.  {\bf One striking
  feature of these figures is the sharp rise of some molecular
  abundances at low velocity. We trace these apparent velocity
  thresholds to temperature barriers or endothermicities of key
  reactions which initiate the production of molecules. Equation
  \eqref{eq:Tmax} allows to relate these critical temperatures to key
  velocities in J-shocks. We list in table (\ref{Flo:bottleneck}) the
  main bottleneck reactions which involve H$_{2}$ and a single atom or
  monoatomic ion along with their characteristic temperatures and J-shock velocities.} At high velocities, the decrease in molecular
  content (around 35 km.s$^{-1}$) corresponds to the absence of
  H$_{2}$ molecules due to dissociation in the shock front.

Figure \ref{Flo:b0.1}(c) illustrates the variation of a few ions of
interest.  The abundance of C$^+$ is controlled mainly by
photo-ionisation of C and charge exchange with PAHs and H$_2$. As the
velocity increases, the post-shock density increases with enhanced charge
exchange, thus decreasing the C$^+$ abundance. For very large
velocities (above 35 km.s$^{-1}$) H$_2$ starts to be dissociated and C$^+$
increases again. In a molecular medium, the abundance of H$_3^+$ is
controlled by the balance between cosmic ray ionisation of H$_2$
(followed by a hydrogen exchange with H$_2$) and recombination of
H$_3^+$. The charge balance links the ionisation degree to the
abundance of C$^+$, thus the abundance of H$_3^+$ is inversely
proportional to C$^+$ with an extra dependence on the square root of
the temperature due to the temperature sensitivity of the recombination
rate.

\begin{table}
\begin{tabular}{ccc}
Reaction & Temperature barrier & Velocity\tabularnewline
\hline
\hline 
O+H$_{2}$ $\rightarrow$OH+ H & 2980 K & 7.5 km.s$^{-1}$\tabularnewline
C$^{+}$+ H$_{2}$ $\rightarrow$CH$^{+}$+ H & 4640 K & 9.4 km.s$^{-1}$\tabularnewline
S+ H$_{2}$ $\rightarrow$SH+ H & 9620 K & 13.5 km.s$^{-1}$\tabularnewline
S$^{+}$+ H$_{2}$ $\rightarrow$SH$^{+}$+ H & 9860 K & 13.6 km.s$^{-1}$\tabularnewline
C+ H$_{2}$ $\rightarrow$CH+ H & 14100 K & 16.3 km.s$^{-1}$\tabularnewline
Si$^{+}$+ H$_{2}$ $\rightarrow$SiH$^{+}$+ H & 14310 K & 16.4 km.s$^{-1}$\tabularnewline
N+ H$_{2}$ $\rightarrow$NH+ H & 14600 K & 16.6 km.s$^{-1}$\tabularnewline
H$_{2}$ dissociation energy & 52000 K & 31.3 km.s$^{-1}$\tabularnewline
\hline
\end{tabular}

\caption{List of bottleneck reactions for molecules formation with their temperature
barrier or endothermicity and the corresponding velocity of the {\bf J-type} shock that is able to
provide such temperature computed using equation \eqref{eq:Tmax}.
 N$^+$ +H$_{2}$ has little or no barrier: 168K.\label{Flo:bottleneck}}

\end{table}

\begin{figure}
(a)\includegraphics[width=8.8cm]{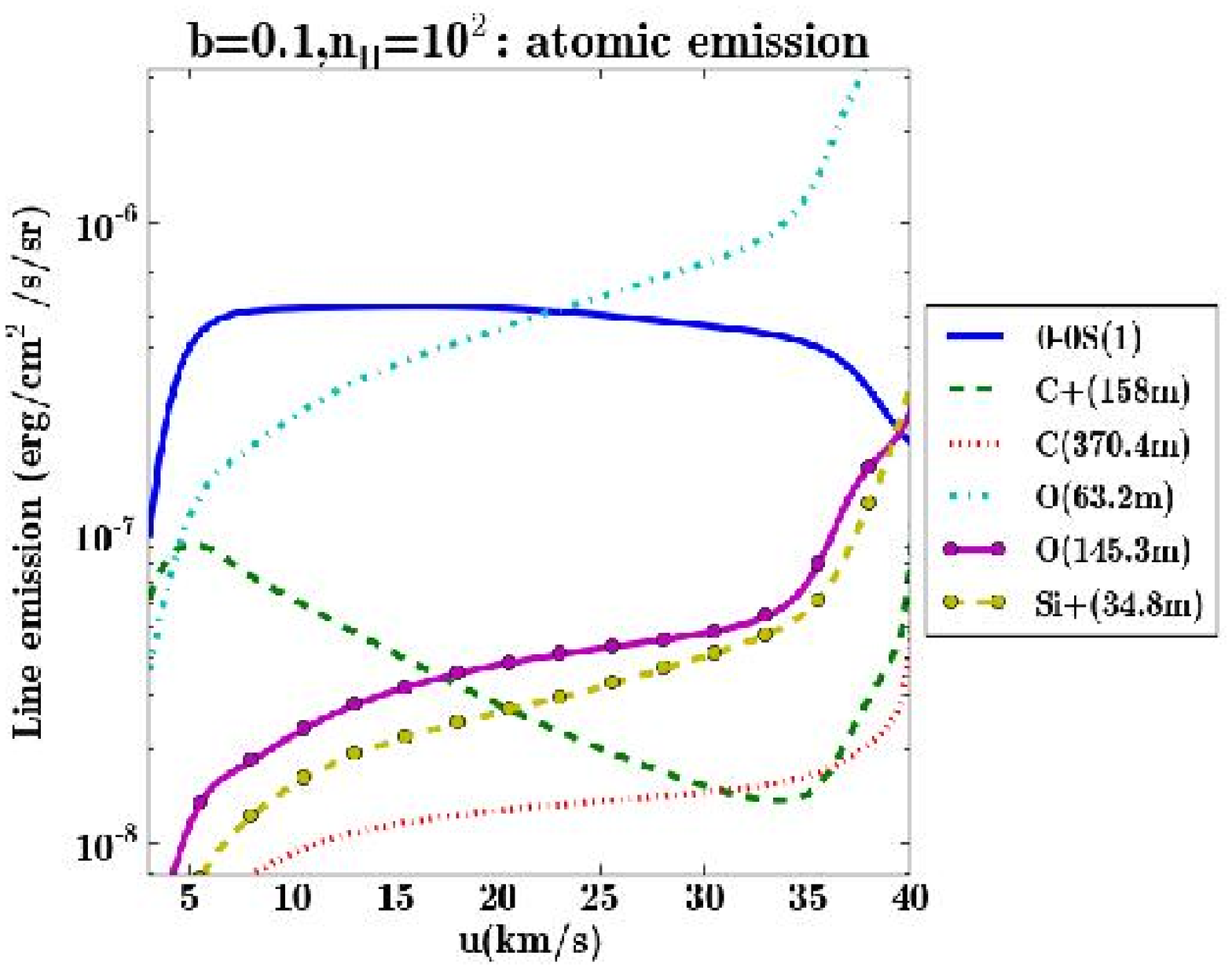}

(b)\includegraphics[width=8.8cm]{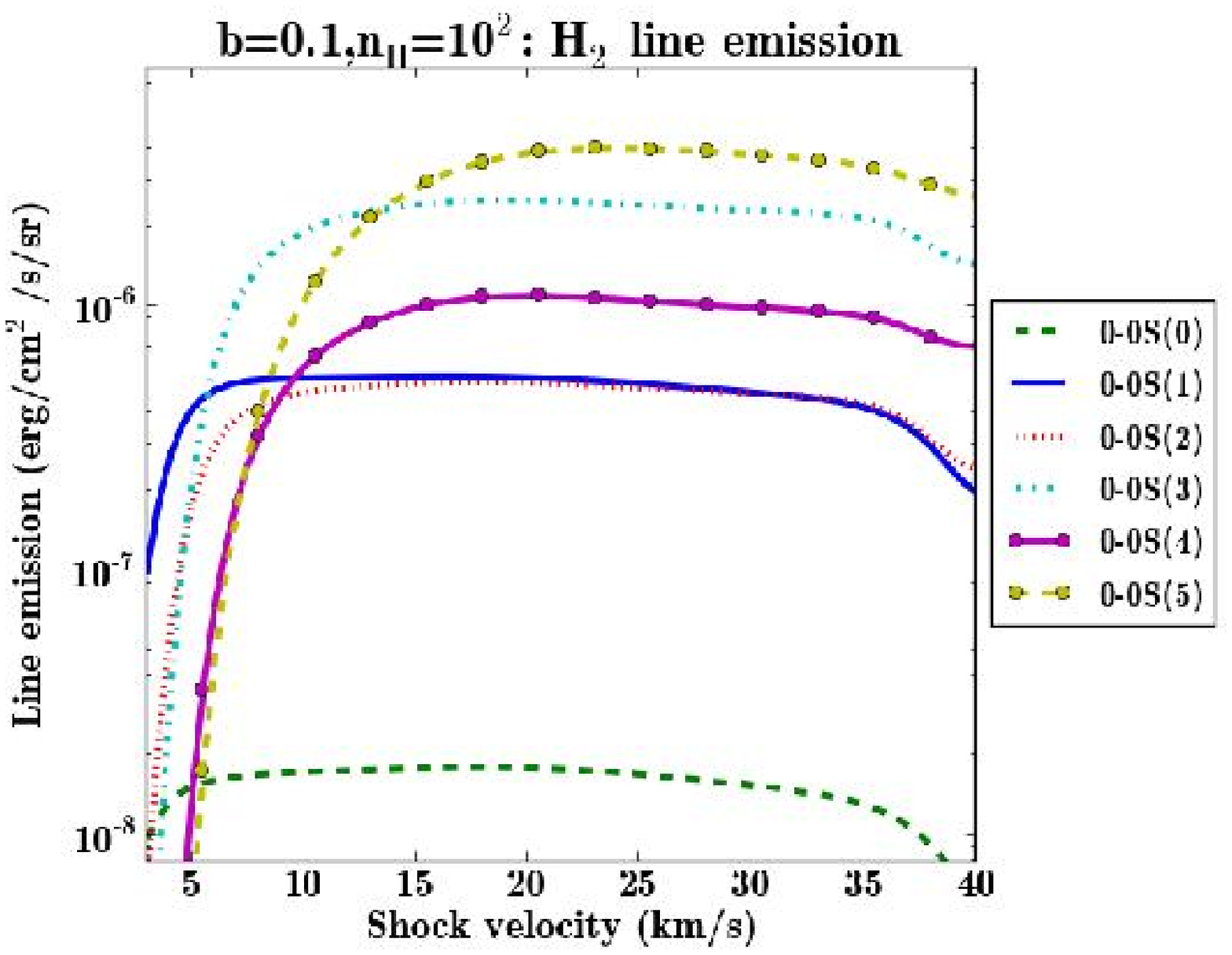}

\caption{\label{Flo:b0.1.lines}(a) Atomic and (b) pure rotational H$_{2}$
line emission for \emph{weakly magnetized} shocks. $n_{\rm H}=10^2~$cm$^{-3}$.}

\end{figure}

On figure \ref{Flo:b0.1.lines} we display the variation of the total
emission of various lines of interest depending on the shock
velocities.  This can be of interest for observers who want to
characterize the velocities of observed shocks.  Figure
\ref{Flo:b0.1.lines}(a) shows some of the most observed atomic lines
with the H$_2$ 0-0S(1) line for comparison.  
The C$^+$ ion emission decrease at moderate velocities is an abundance effect. 
Atomic O cooling is enhanced at moderate velocities because
these shocks dissociate H$_2$: this slows down the cooling and yields a
temperature plateau where O is the dominant cooling agent. The
stronger the shock, the more time it takes for H$_2$ to reform, then the
plateau widens and the O emission gets larger. 

 Figure
\ref{Flo:b0.1.lines}(b) shows how the lowest energy lines of H$_2$
change with the velocity of the shock. The relative strengths of these
lines vary greatly between $u$=3 km.s$^{-1}$ and $u$=10 km.s$^{-1}$. For example, the
ratio S(1)/S(3) decreases from more than a hundred at 3 km.s$^{-1}$ to about
a fifth at 10 km.s$^{-1}$.  Indeed, the shock peak temperature in this range
of velocities spans the energies of the upper levels of these
transitions ($J=7$ corresponds to 3474 K, for instance, and see figure
\ref{Flo:b0.1.T}(a) for how the maximum temperature in the shock varies with shock velocity): the
excitation diagram in the low levels of H$_2$ is greatly affected
in this range of temperatures.  By contrast, the aspect of the
excitation diagram does not change above 10 km.s$^{-1}$: the emissivities of
these lines are unable to discriminate the shock velocity $u$ for $u>10~$km.s$^{-1}$. Indeed, above 10 km.s$^{-1}$ the temperature
experienced in these shocks is much greater than the $J=7$
energy. Observers should use lines with upper levels with higher $J$
values (such as rovibrational lines) to probe the velocities of these shocks or they should use
diagnostics based on atomic lines.

\begin{figure}
\includegraphics[width=8.8cm]{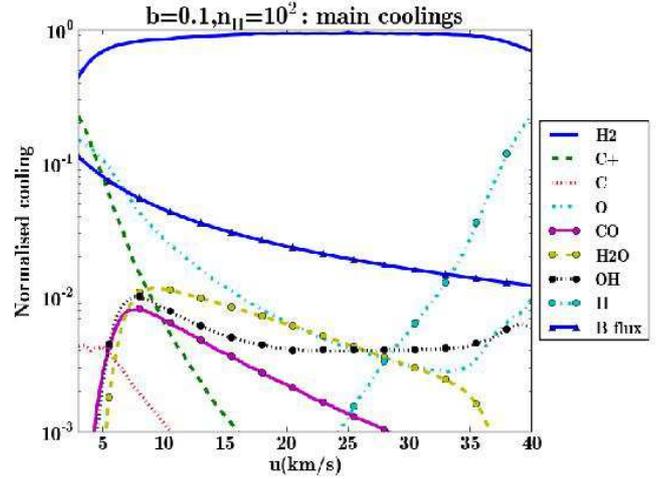}
\caption{Cooling integrated along the shock length, normalised by kinetic power $\frac{1}{2}\rho u^{3}$ in \emph{weakly
magnetized} shocks, for $n_{\rm H}=10^{2}\mbox{cm}^{-3}$.
The solid line with triangles shows the fraction of the power which is transferred
into a flux of magnetic energy. 
On this figure, H$_2$ cooling incorporates the cooling from 200 lines between the 
149 H$_2$ levels included in the simulation, whereas figure \ref{Flo:b0.1.lines} considers 
only 6 individual lines amongst the lowest levels of H$_2$. \label{Flo:b0.1.cool}}

\end{figure}

On figure \ref{Flo:b0.1.cool} we examine what fraction
of the kinetic energy flux input 
\begin{equation}
  I_{\mbox{kinetic}}=\frac12 \rho u^3
\end{equation}
into the shock is radiated away by
each coolant. Individual coolings are integrated through the whole shock structure as follows:
\begin{equation}
  I_{\mbox{cool}}=\int_{x_0}^{x_1}\Lambda \mbox{d}z 
\end{equation}
where $\Lambda$ is the local rate of cooling and $x_1$ is the end point of the shock (ie: where the temperature decreases back to 20\% above the pre-shock temperature). {\bf We also display the magnetic energy flux accross the shock 
\begin{equation}
I_{\mbox{B flux}}=\frac{B_e^2}{4\pi}u_i
\end{equation}
where $u_i$ is the ion velocity and $B_e$ is the magnetic field at the end of the shock.}
 H$_{2}$ cooling dominates
almost everywhere except: (i) at very low velocity, where atomic O or C$^+$ cooling
takes over depending on the density, and (ii) above the velocity for H$_{2}$ dissociation (which
is lower for higher densities) where atomic O and H (Lyman-$\alpha$)
cooling take over. At density $n_{\rm H}=10^{4}\mbox{ cm}^{-3}$, and
below 20 km.s$^{-1}$, note that H$_{2}$O cooling becomes important.

\subsection{Strongly magnetised shocks (b=1)}
\label{modelsC}

\begin{figure}
(a)\includegraphics[width=8.8cm]{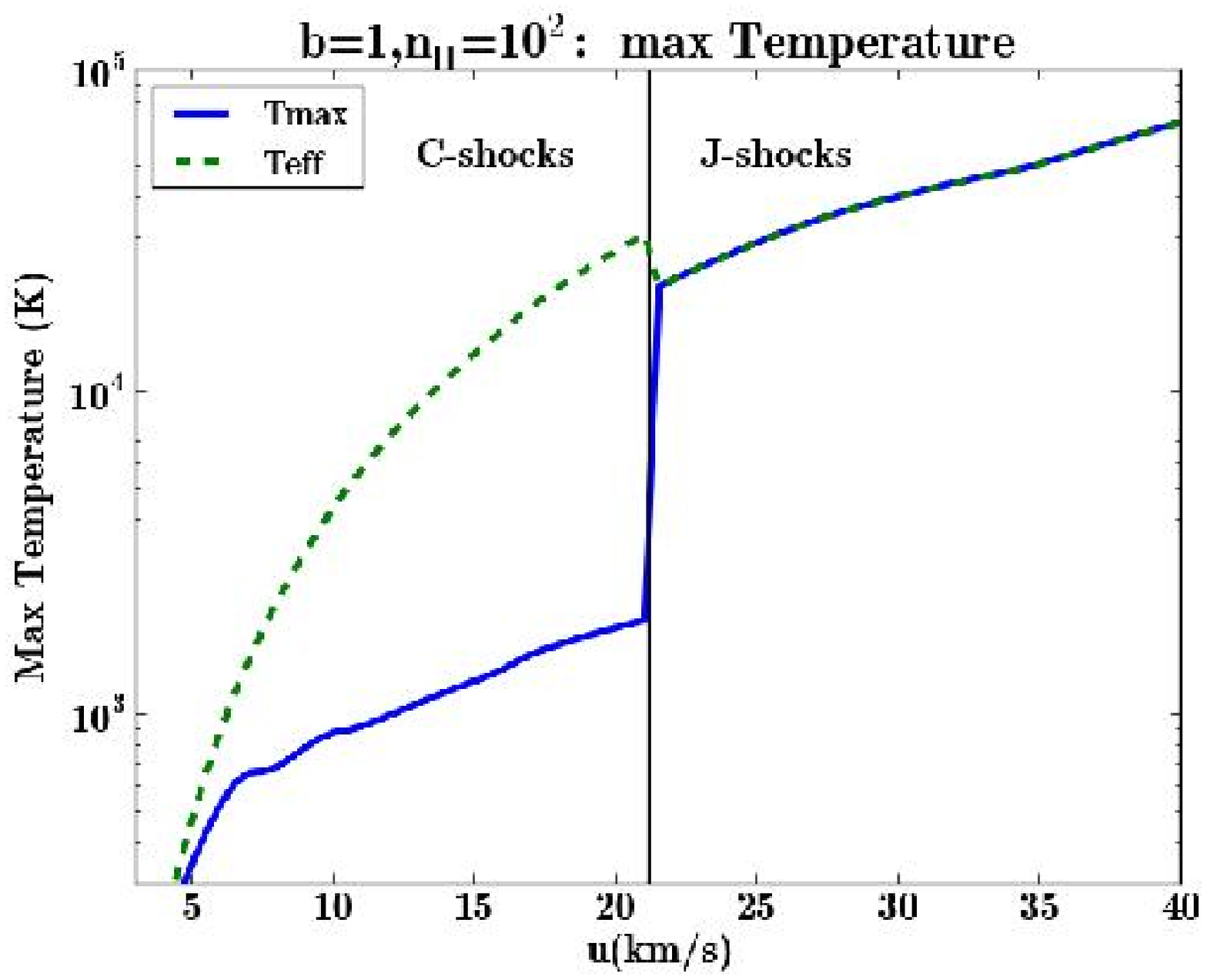}

(b)\includegraphics[width=8.8cm]{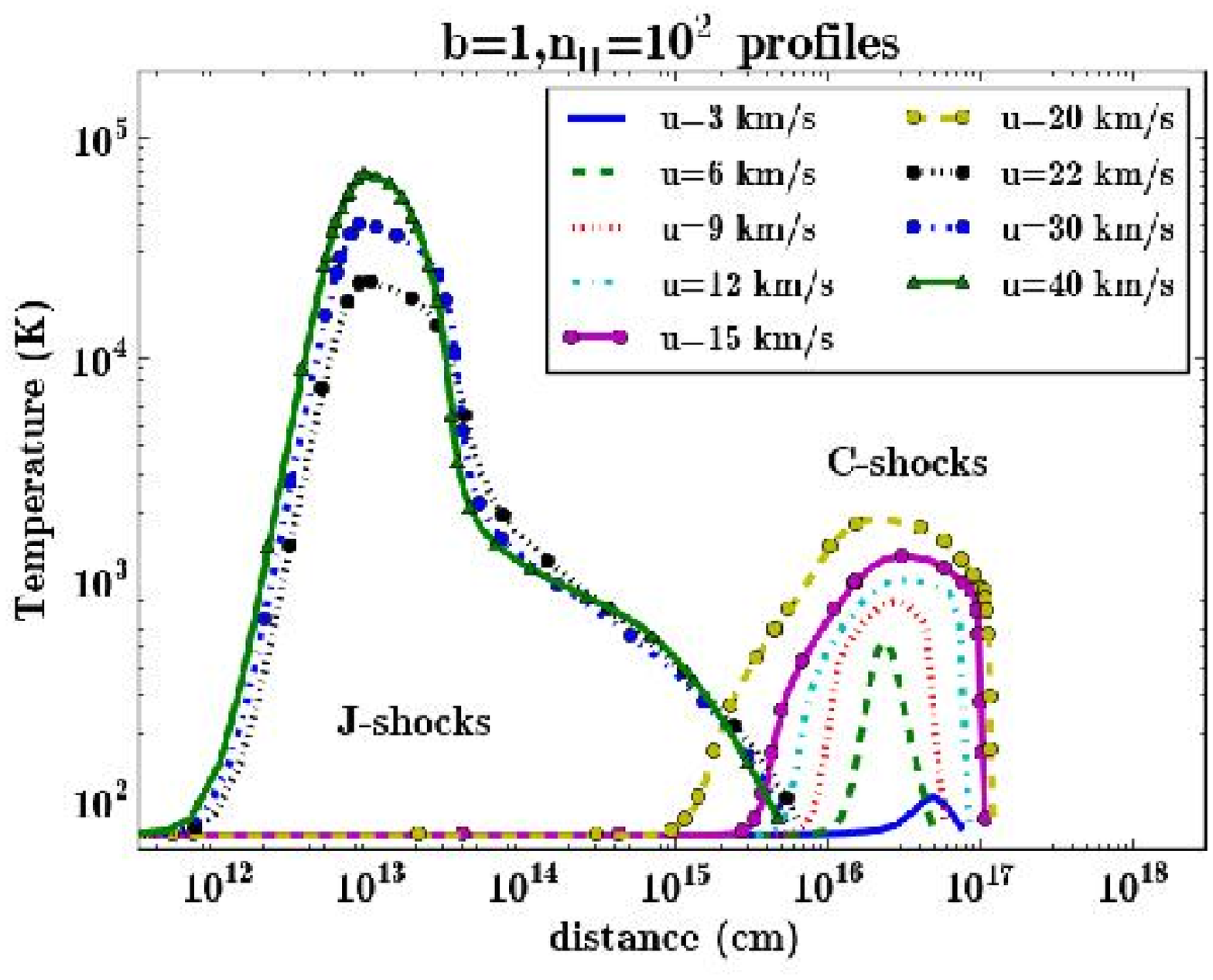}

\caption{(a) Maximum temperature in the \emph{highly magnetized} shocks (b)
Temperature profiles for some representative \emph{highly magnetized}
shocks{\bf, the fluid flows from left to right with the pre-shock on the left and the post-shock on the right}. $n_{\rm H}=10^2~$cm$^{-3}$.\label{Flo:b1.T}}

\end{figure}

In a C-shock, the kinetic energy is continuously transformed into
thermal energy via ion-neutral friction. As a result, the heating
is spread out on a much larger region than for J-shocks and the peak temperature
is hence much smaller as seen on figure \ref{Flo:b1.T}(a). However,
reactions between neutral and ion species benefit from the ion-neutral
drift, and the effective temperature which is used to compute the
reaction rate is higher (see \citealp{1986MNRAS.220..801P}). For
instance, the effective temperature for reactions involving C$^{+}$
and H$_{2}$ is\begin{equation}
T_{\mbox{eff}}(\mbox{C}^{+},\mbox{ H}_{2})\simeq\frac{2T_{n}+12T_{i}}{2+12}\,+\,\frac{2\times12}{2+12}\,\frac{m_{p}}{3\, k_{B}}(u_{n}-u_{i})^{2}\end{equation}
where $(u_{n}-u_{i})$ is the local drift velocity between the ions
and the neutrals, 
$m_p$ is the proton mass and $T_{n}$ and $T_{i}$ are respectively the temperatures
of the neutral and ionised gas. Figure \ref{Flo:b1.T}(a) shows 
the maximum effective temperature of the reaction C$^{+}$ + H$_{2}$, which
 incidentally is very close to the maximum temperature which would
be obtained in the corresponding J-shock. 

Compared to the temperature profile of a J-shock, the temperature
profile of a C-shock is broad, with a lower temperature: see figure
\ref{Flo:b1.T}(b). However, within each type of shock, the total
size of the shock structure varies over less than an order of magnitude
for the range of velocities we tested, as is apparent on figure \ref{Flo:b1.T}(b). This is even more striking
for the flowing time scales across the shock structures which, for $n_{\rm H}=10^2~$cm$^{-3}$, varies
only from 8000 to 6000 years for the C-shocks and from 1500 to 1000
years for the J-shocks. As mentioned above, this holds for higher densities as well.

\begin{figure}
(a)\includegraphics[width=8.8cm]{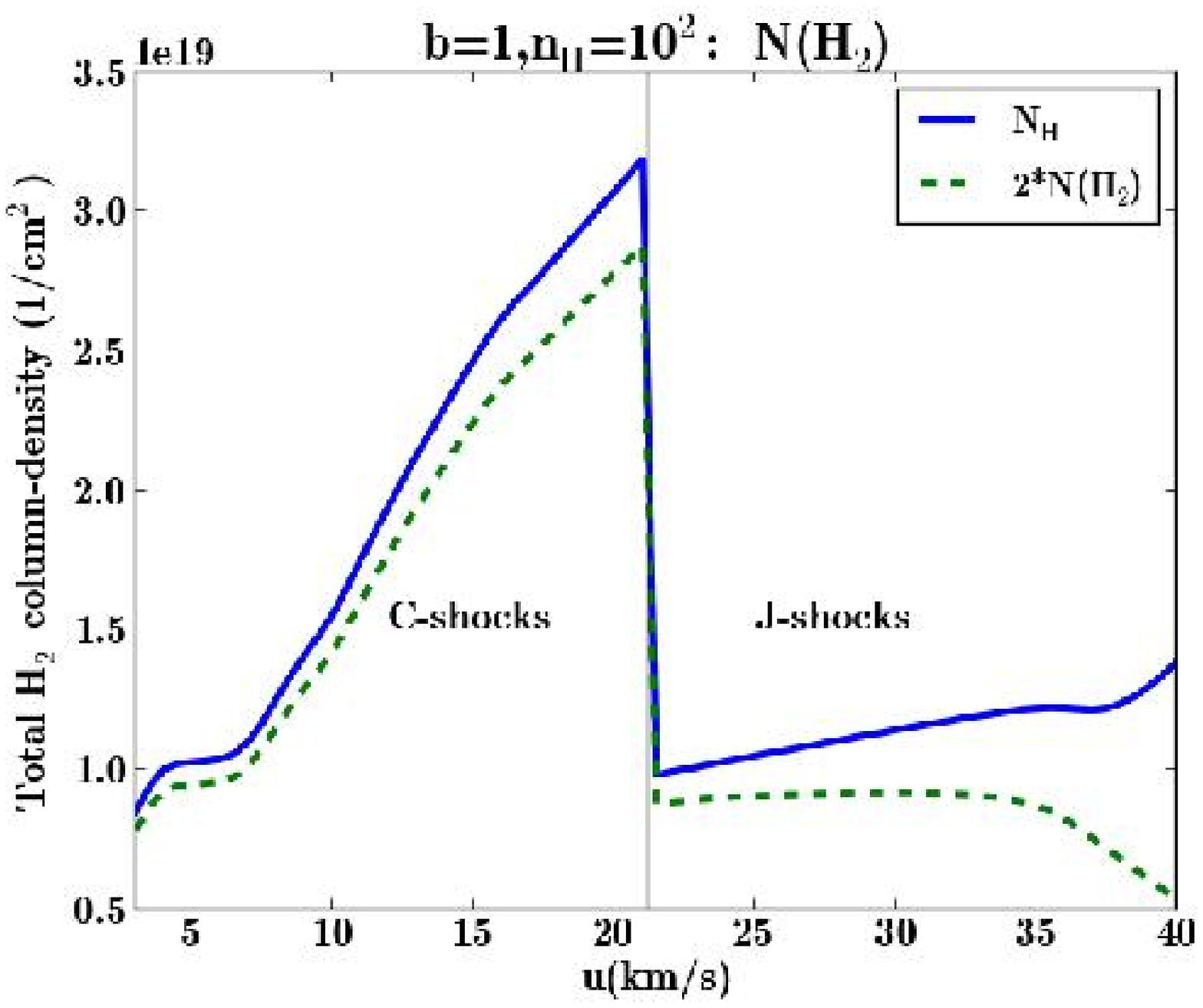}

(b)\includegraphics[width=8.8cm]{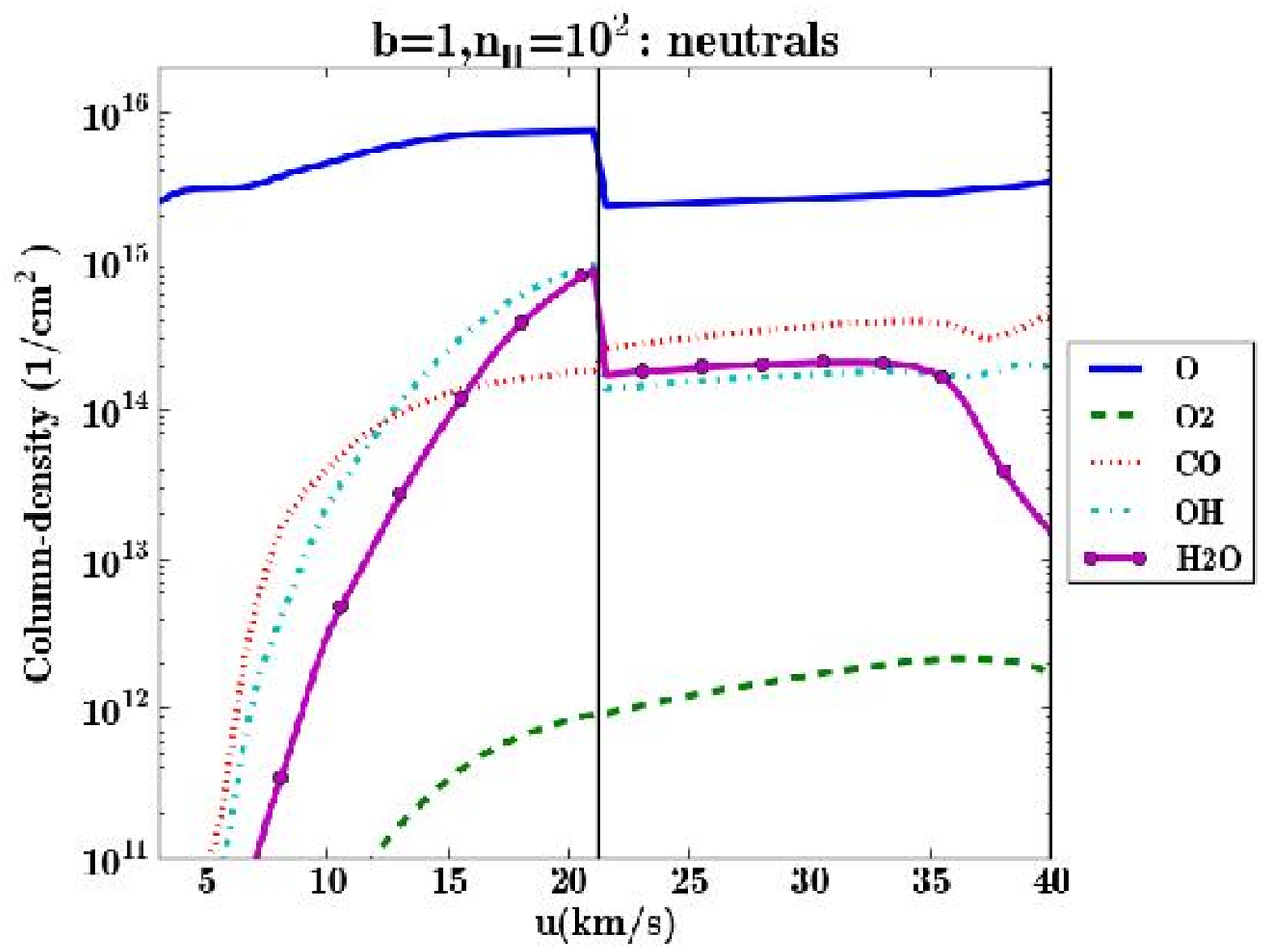}

(c)\includegraphics[width=8.8cm]{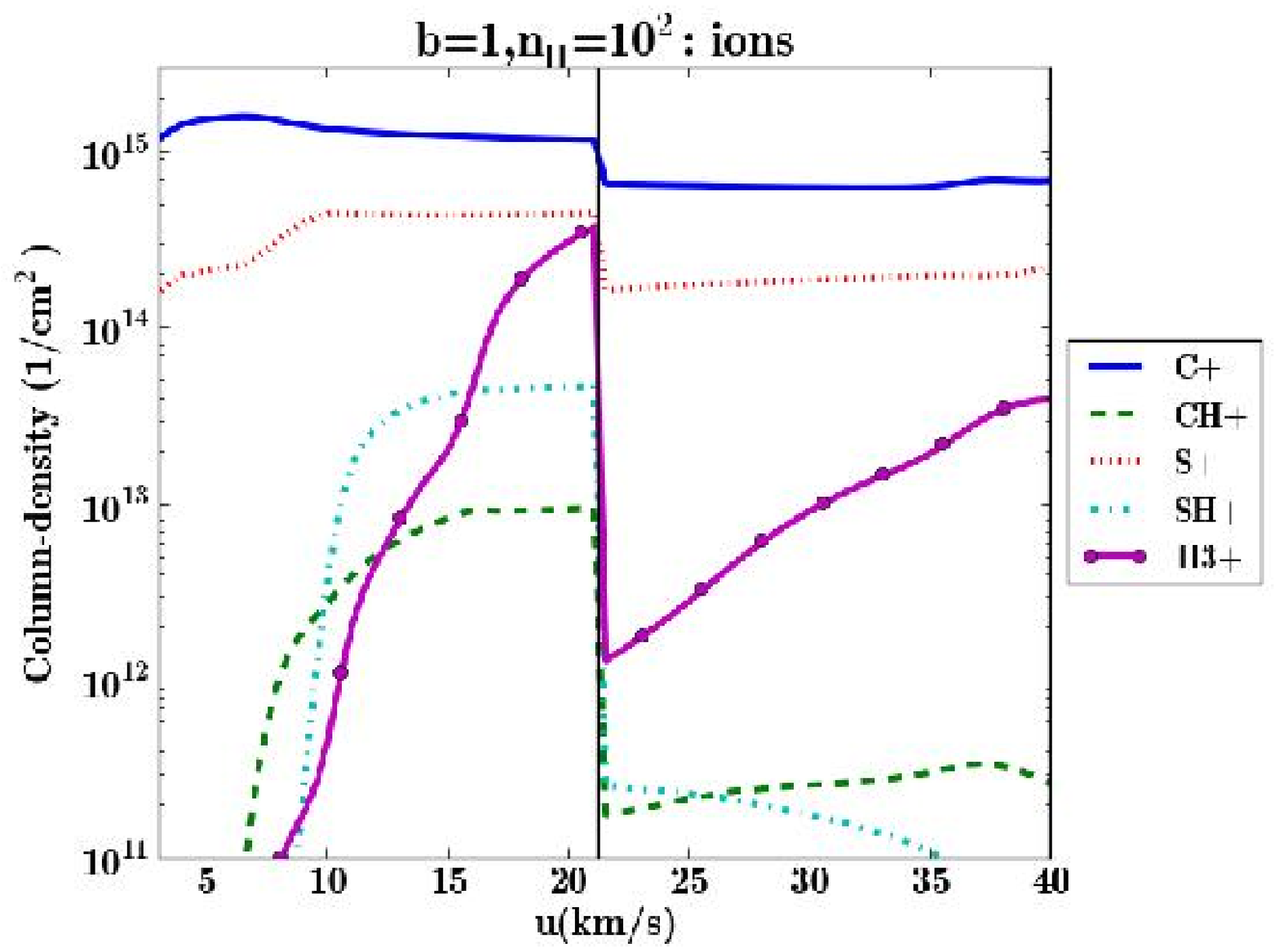}

\caption{Same as figure \ref{Flo:b0.1} for \emph{highly magnetized }shocks ($b=1$). $n_{\rm H}=10^2~$cm$^{-3}$. \label{Flo:b1}}

\end{figure}

 Slightly above 20 km.s$^{-1}$, all shocks in our grids of models are
J-shocks, but  the strength of the magnetic field varies from $b=0.1$ in  the previous subsection to  $b=1$ in this subsection.
 The higher magnetic field limits the
compression in the shock and the collisional processes take longer to
occur. In particular, the cooling length for the J-shocks above 20 km.s$^{-1}$ is between $4\times10^{15}~$cm and $7\times10^{15}~$cm: much wider than for the corresponding weakly magnetised J-shocks.
 The lower density but larger size of the shock impacts on
the chemical composition and structure in subtle ways.  A careful comparison of the right hand
sides of figures \ref{Flo:b0.1} and \ref{Flo:b1} shows the column-density of H nuclei is
higher in magnetised shocks and  moderate variations
in the chemical composition are noticeable.

Figure \ref{Flo:b1} shows the molecule production over the range
of models in our grid at $n_{\rm H}=10^2~{\rm cm}^{-3}$. The transition from C-shocks (velocity lower
than 21 km.s$^{-1}$) to J-shocks (higher shock speed) is clear. The column-densities of most species have a drop at  21 km.s$^{-1}$ which expresses a boosted molecular production in C-shocks.
Three effects are at play which favour molecules production in C-shocks.

First, the ion-neutral drift helps to overcome reaction barriers of
ion-neutral reactions (which impact C-bearing species, for example, whose
bottleneck reaction is C$^+$+H$_2$).
Second, the resulting frictional heating keeps the temperature warm throughout the shock. 
By comparison a J-shock is very hot at the peak temperature right behind the viscous front 
but quickly cools down in the trailing relaxation layer. Third, the steady frictional heating slows down
cooling and compression in the relaxation layer and the net result is an increased total 
$N_{\rm H}$ column-density in C-shocks compared to J-shocks (see the drop at 21 km.s$^{-1}$ in figure \ref{Flo:b1}a). This is the main factor which impacts those molecular species whose
production relies mainly on  neutral-neutral reactions
(such as O-bearing species). In particular, these species show a more gradual rise at low
 velocities, because of the slower rise of the maximum
neutral temperature in C-shocks. 

\begin{figure}
(a)\includegraphics[width=8.8cm]{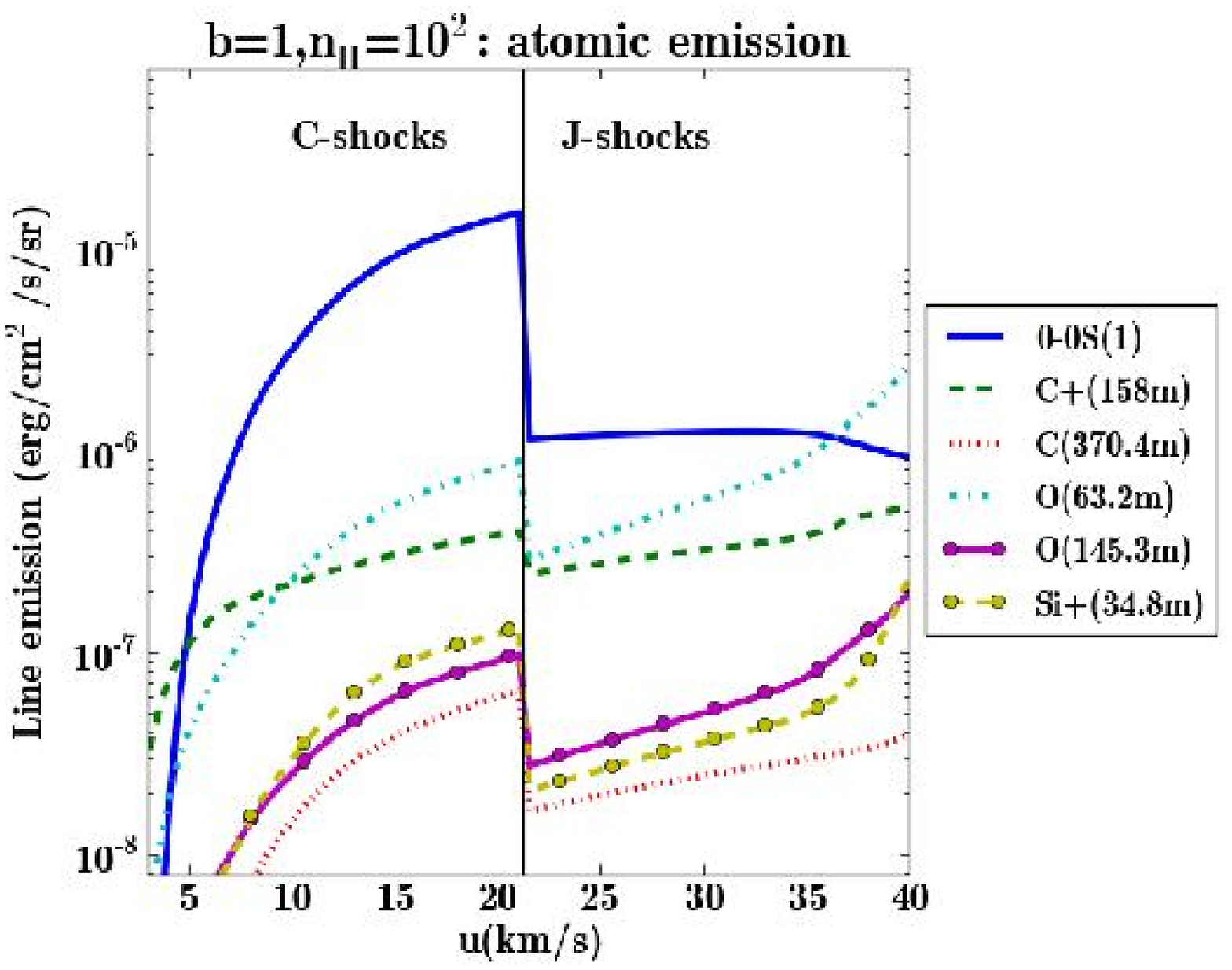}

(b)\includegraphics[width=8.8cm]{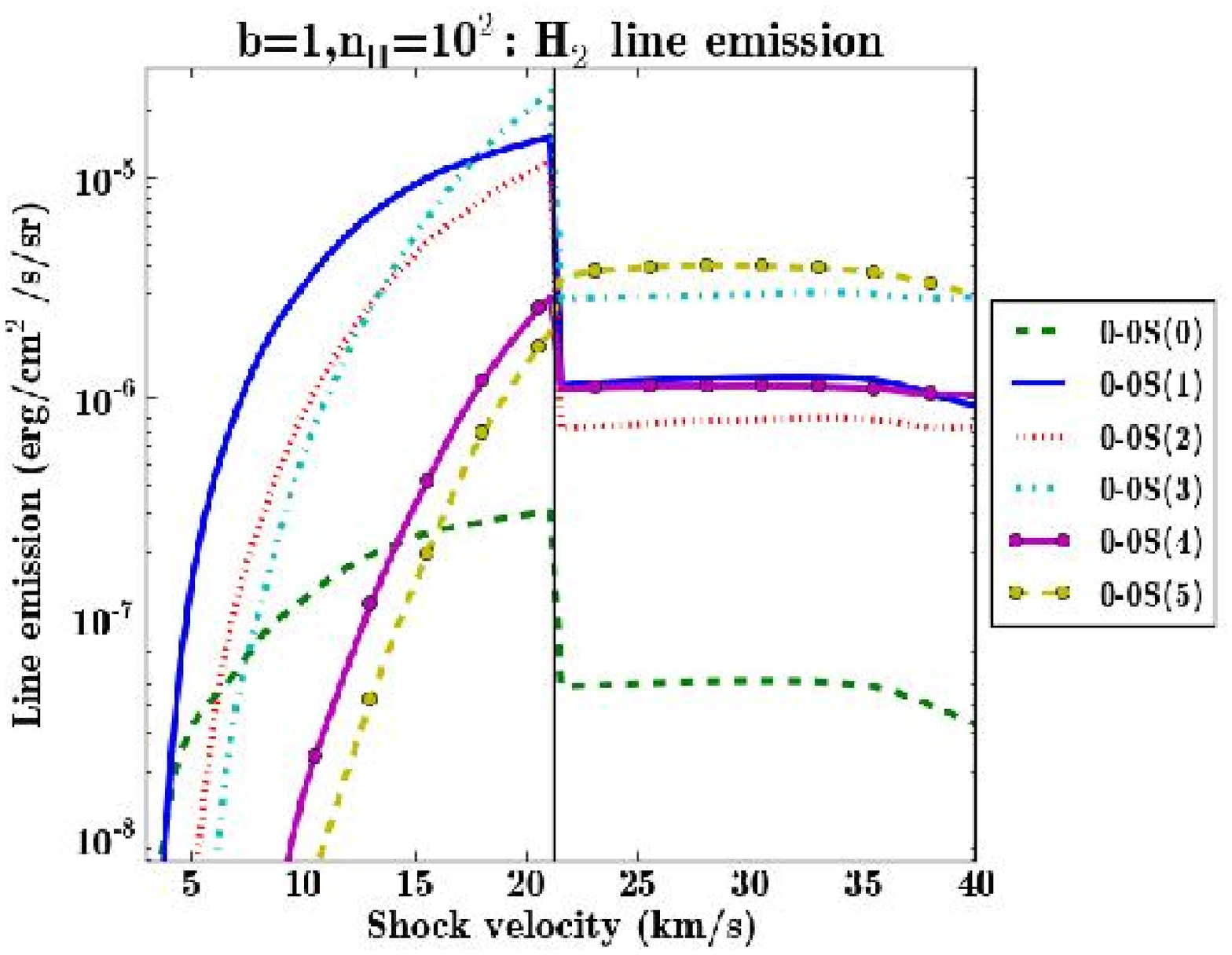}

\caption{(a) Atomic and (b) rotational H$_{2}$ lines emission for \emph{highly
magnetized} shocks. $n_{\rm H}=10^2~$cm$^{-3}$.\label{Flo:b1.atomic_lines}}

\end{figure}

We display the total integrated emission in various lines of interest
on figures \ref{Flo:b1.atomic_lines}(a) and
\ref{Flo:b1.atomic_lines}(b): as for figure \ref{Flo:b0.1.lines},
these give clues on which observations constrain what shock velocity.
In particular, the emissivities of H$_2$ lines with low J upper levels now
vary from 3 to 20 km.s$^{-1}$, but are independent of velocity above 20 km.s$^{-1}$.
The neutral temperature in C-shocks rises more slowly than J-shocks as
shown by the comparison between figures \ref{Flo:b0.1.T} and
\ref{Flo:b1.T}, but the maximum temperature in C-shocks is more
representative of the temperature in the whole mass of the shock and there is more column-density in C-shocks. As a
result, the left hand sides (between 3 and 20 km.s$^{-1}$) of figures
\ref{Flo:b1.atomic_lines}(a) and \ref{Flo:b1.atomic_lines}(b) look
like blow ups of figures \ref{Flo:b0.1.lines}(a) and
\ref{Flo:b0.1.lines}(b) shifted to greater emissivities. However,
C$^+$ stands out even more for C-shocks because its excitation
benefits from the ion-neutral drift. The right hand sides of figures
\ref{Flo:b1.atomic_lines}(a) and \ref{Flo:b1.atomic_lines}(b) are
similar to the corresponding weakly magnetised J-shocks seen in
figures \ref{Flo:b0.1.lines}(a) and \ref{Flo:b0.1.lines}(b): the
emissivities are poorly affected by the density change due to the
magnetic field except for C$^+$.

\begin{figure}
\includegraphics[width=8.8cm]{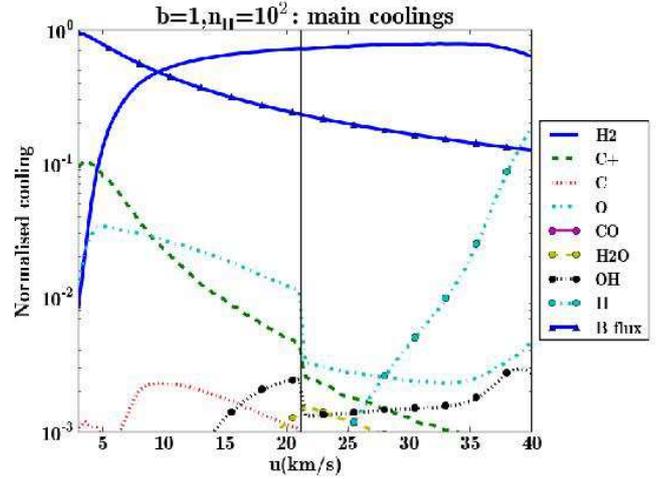}

\caption{Cooling normalised by kinetic power in \emph{highly magnetized} shocks
for pre-shock density $n_{\rm H}=10^{2}\mbox{cm}^{-3}$.
The solid line with triangles shows the fraction of the power which is transferred
into a flux of magnetic energy.\label{Flo:b1.cool}}

\end{figure}

In figure \ref{Flo:b1.cool} we examine,
averaged over the whole shock structure as in figure \ref{Flo:b0.1.cool}, what fraction of the kinetic
energy flux input into the shock is radiated away by each coolant in the low 
density case ($n_{\rm H}=10^{2}\mbox{ cm}^{-3}$). At very
low velocity, C$^{+}$ is the most efficient coolant (rather than
O for weakly magnetised shocks), but H$_{2}$ takes over for shock
velocities shortly above 5 km.s$^{-1}$. Molecules such as CO, H$_{2}$O or
OH carry less than a percent of this energy flux. The kinetic energy flux is mainly
transferred into a magnetic energy flux (via field compression) for velocities
below 10 km.s$^{-1}$ and is mainly radiated away above that velocity. At higher densities, the low velocity coolant becomes O. 
It is also interesting to note that at $n_{\rm H}=10^{4}$
cm$^{-3}$ H$_{2}$O becomes an important cooling agent for low
velocities. \citet{2010MNRAS.406.1745F} also quoted the fraction
of energy radiated away in various cooling processes in their shock
models. However, in the present work the irradiation field photo-dissociates
CO and ionises C so that C$^{+}$ cooling is more prominent and CO
cooling is less important than in their work, but the results are
otherwise similar. In particular, they also find that the magnetic energy flux
stores 50\% of the injected power at $u=10$ km.s$^{-1}$.

\section{PDF of shocks }

\subsection{Method of fit}

Numerical simulations of driven supersonic turbulence show that the medium experiences a
whole range of shock speeds: \citet{2000A&A...362..333S} show that
the probability distribution function (PDF) of the velocity jumps
follows a decreasing power law with an exponential cut off at a Mach
number of a few. \citet{2000A&A...356..279P} carefully separated
vortical from compressional contributions in the velocity field of
a compressible simulation of decaying supersonic turbulence: they
uncover an exponential distribution for the convergence time scales
$-1/\mbox{div}(v)$ where $v$ is the fluid velocity. Subsonic wind
tunnel experiments display an exponential distribution of velocity
increments (\citealt{2008PhFl...20c5108M}). 

These results hint at a variety of possible PDF shapes for the statistics
of shocks, all favourably weighted towards low velocity shocks. {\bf \citet{2009A&A...502..515G} 
have modeled Spitzer H$_2$ observations of SQ with only two discrete values of the shock velocities}. Here, we aim at fitting
a statistical distribution of shocks to a collection of observed quantities.
Guided by the above remarks on turbulence, we fit various shapes of
shocks PDF: decaying exponentials, power-laws, Gaussians
and a piece-wise exponential fit. 

We consider among $j=1..M$ observable quantities an observable $X_{j}(u)$
associated to a steady shock of velocity $u$ (expressed
in km.s$^{-1}$). We note $f(u,\{p_{i}\}_{i=1..N})$ the probability density
parametrised by the set of parameters $\{p_{i}\}_{i=1..N}$ . Of course,
the fitting makes sense only as long as the number of observed quantities
$M$ is sufficiently greater than the number of fitted parameters
$N$. For example, an exponential PDF reads \begin{equation}
f_{e}(u,p_{1})=C(p_{1})\,\exp(-p_{1}u)\end{equation}
 with the normalisation constant \begin{equation}
C(p_{1})=1/\sum_{u\geqslant3}^{u\leqslant40}\,\exp(-p_{1}u).\end{equation}

We then compute observable quantities averaged over the PDF of shocks
as\begin{equation}
\bar{X}_{j}(\{p_{i}\}_{i=1..N})=\sum_{u\geqslant3}^{u\leqslant40}X(u)\, f(u,\{p_{i}\}_{i=1,N}).\end{equation}

In order to ease the fitting process, we normalise the above values
to the sum of the observable quantities:\begin{equation}
\hat{X_{j}}=\frac{\sum_{j=1..M}X_{j}^{0}}{\sum_{j=1..M}\bar{X_{j}}}\bar{X_{j}}\end{equation}
where $\{X_{j}^{0}\}_{j=1..M}$ are the actual observed values. Note
that this normalisation renders irrelevant the normalisation constant
of each PDF, but it decreases by one unit the number of degrees of freedom $M_\mathrm{free}=M-N-1$. We finally find the optimal set of parameters $\{p_{i}^{0}\}_{i=1..N}$
which minimises the quantity\[
\chi^{2}(\{p_{i}\}_{i=1..N})=\frac{1}{M_{\mbox{free}}}\sum_{j=1..M}\left(\log_{10}\hat{X_{j}}-\log_{10}X_{j}^{0}\right)^{2}/W_{j}^{2}\]
where $W_{j}$ is the uncertainty on the observed value $\log_{10}X_{j}^{0}$.
We also retrieve estimates $\{E_{i}\}_{i=1..N}$ of the errors on
the parameters by computing the diagonal elements of the covariance
matrix of $\frac{\partial^{2}}{\partial p_{i}^{2}}\chi^{2}$ on the
optimal set.

Table \ref{Flo:PDF} summarises the various PDFs we tried and their
parametrization. The exponential and power-law PDFs are motivated by
the above mentioned work on the statistical distribution of shocks
in numerical and laboratory experiments. The piece-wise exponential
PDF is built on intervals designed such that the observable quantities
vary significantly (cf. figures \ref{Flo:b0.1.lines} and \ref{Flo:b1.atomic_lines}).
Otherwise, the parameters would be degenerate. We use as parameters
the PDF value at the transition between these intervals. Thus, these
parameters' optimal values and corresponding error bars directly probe the PDF and its uncertainty
at these chosen locations. Finally, we use a single or a double Gaussian PDF to
mimic the fit of one shock or two independent shocks: this allows to connect with
the work of \citet{2009A&A...502..515G}.

\begin{table}
\begin{tabular}{ccc}
\hline 
PDF & formula & N\tabularnewline
\hline
\hline 
Power-Law & $u^{-p_{1}}$ & 1\tabularnewline
Exponential & $\exp(-p_{1}u)$ & 1\tabularnewline
Piece-wise exponential & \begin{tabular}{c}
at $u=3,\,10,\,20,\,40$\tabularnewline
$f(u)=1,\, p_{1},\, p_{2},\, p_{3}$\tabularnewline
\end{tabular} & 3\tabularnewline
1-Gaussian & $e^{-(u-p_{1})^{2}}$ & 1\tabularnewline
2-Gaussian & $e^{-(u-p_{1})^{2}}+p_{3}e^{-(u-p_{2})^{2}}$ & 3\tabularnewline
\hline
\end{tabular}

\caption{Template PDFs adjusted to the observed data (normalisation
constants are discarded as irrelevant, see text). \label{Flo:PDF}}

\end{table}

In addition, we note that the choice of the boundaries (upper and lower
velocities) for the range of our grid of models does not affect significantly our
results. This stems from the fact that the shock emission is very
small at low velocity and high velocity shocks are very rare in the
solutions we obtain. Finally, grids of models with velocity spacings
of 1 km.s$^{-1}$ and 0.5 km.s$^{-1}$ gave similar results, which validates the resolution
we chose.

\subsection{Applications}

\subsubsection{The Stephan's Quintet galaxy collision}

In this section we use our model results to fit H$_2$ line emission
from the galaxy-wide shock in SQ. This shock structure, first
identified in radio continuum observations and X-ray images, was
discovered to be a luminous H$_2$ source with {\it Spitzer}
\citep{2006ApJ...639L..51A, 2010ApJ...710..248C}. It is associated
with the entry of a galaxy into a group of interacting galaxies with a
relative velocity $\rm \sim 1000 \,$ km.s$^{-1}$. The luminosity of
the molecular gas in H$_2$ rotational lines is observed to be larger
than that of the plasma in X-rays.  \citet{2009A&A...502..515G}
present a first interpretation where the H$_2$ emission is powered by
dissipation of turbulence driven by the large scale collision.
CO observations have since shown that the kinetic energy of the molecular
gas is larger than the thermal
energy of the hot (X-ray emitting) plasma. It is the main energy
reservoir available to power the H$_2$ line emission 
\citep{2012ApJ...749..158G}. \citet{2009A&A...502..515G} show that the
H$_2$ excitation is well fit by a combination of two MHD shocks with
velocities of 5 and $20 \, $km.s$^{-1}$ in dense gas ($n_{\rm H} \sim 10^3$ to
$ 10^4 \, $cm$^{-3}$), using the shock models of
\citet{2003MNRAS.343..390F}, which do not take into account UV
radiation.  Our grid of models allows us to test an alternative
interpretation where the H$_2$ emission is accounted for with MHD
shocks in lower density UV irradiated gas.  In doing so we
quantify a solution where the physical state of the H$_2$ gas in SQ is
akin to that of the cold neutral medium in the Galaxy.
  Evidence that the gas is magnetized comes
from radio continuum observations.  The synchrotron brightness of the
shock yields a magnetic field value $B\sim 10~\mu G$, assuming
equipartition of magnetic and cosmic-ray energy (Xu et al. 2003).
This value is comparable to the field strength reported by
\citet{2010ApJ...725..466C} for the Galactic diffuse ISM. The mean UV
radiation field in the SQ shock estimated to be $G_{0}=1.4$
(\citealp{2010A&A...518A..59G}), is also close to the reference value
for the ISM in the Solar Neighborhood.

The wide extension (30~kpc) of the SQ shock suggests that in the
relaxation layer which follows the main high velocity ($\rm \sim 600$
km.s$^{-1}$) shock, a large number of sub-structures are formed which
then collide and yield a range of sub-shocks with much smaller
velocities (see \citealt{2009A&A...502..515G}). Simulations of large
scale shocks subject to thermal instability indeed show that
turbulence is sustained in the trailing relaxation layer
(\citealt{2002ApJ...564L..97K} and later
\citealt{2005A&A...433....1A,2007A&A...465..431H,2010A&A...511A..76A}).
This connects with the theoretical work on turbulence mentioned above,
and justifies our investigation with continuous PDFs of shock
velocities.  The observable targets which we aim at reproducing are
the emission values for the six H$_{2}$ lines 0-0 $S(0)$ to $S(5)$ in table 1 of
\citet{2010ApJ...710..248C} which are integrated over the main shock
structure.

\begin{figure}
\includegraphics[width=8.8cm]{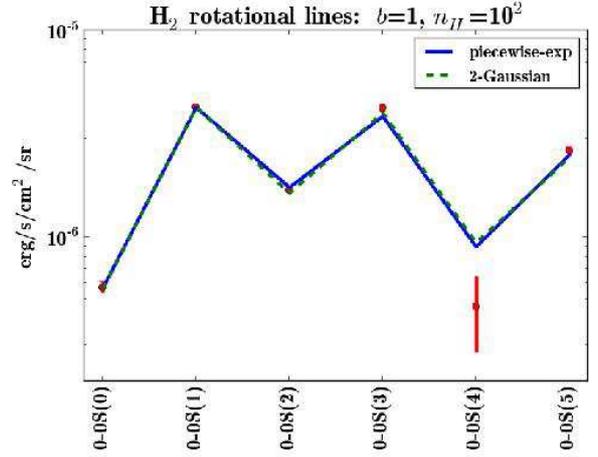}
\caption{Observed fluxes of the H$_2$ lines in the SQ (red dots with error bars: \citealp{2010ApJ...710..248C},
see also table \ref{Flo:results}) with the results of our best two
models. {\bf Note most error bars are so small they remain barely visible.} \label{Flo:fit_fig1}}

\end{figure}

Table \ref{Flo:chi2} shows the best $\chi^2$ obtained in all our
attempts to fit the data. Two solutions stand out with $\chi^2$ values
reasonably close to one, namely: the piece-wise exponential and the two
Gaussians at $b=1$ and $n_{\rm H}=10^{2}~\mbox{cm}^{-3}$ (their
$\chi^2$ values are emphasised in bold faces in table
\ref{Flo:chi2}). 
Figure \ref{Flo:fit_fig1} illustrates the quality of the fit on the
observed H$_{2}$-lines. Our models reproduce quite accurately every
H$_{2}$-line except for 0-0 $S(4)$\footnote{This line actually lies at the position of the 7.7$\mu m$
PAH emission feature, which makes it difficult to measure accurately.}.  For comparison, when we use the same technique and same grid of
models as \citet{2009A&A...502..515G} (ie: models without irradiation at
$n_{\rm H}=10^{4}~$cm$^{-3}$), we find the best fit is obtained for
 two shocks of velocity 5
and 22 km.s$^{-1}$ in proportion 1:0.008 with a reduced $\chi^2=78$ 
(compared to $\chi^2=15.8$ for velocities 3.7 and 21 km.s$^{-1}$ 
in proportion 1:0.0002 in our 2-Gaussians high-density case).
The coincidence of the optimal parameters with our 2-Gaussian solution
is surprising, but the improvement on the $\chi^2$ in the present work
is also striking. Irradiation does really improve the comparison with
the observations. Our work also shows that low-density solutions are even
more viable than the previously found high-density solutions. 

\begin{figure}
\includegraphics[width=8.8cm]{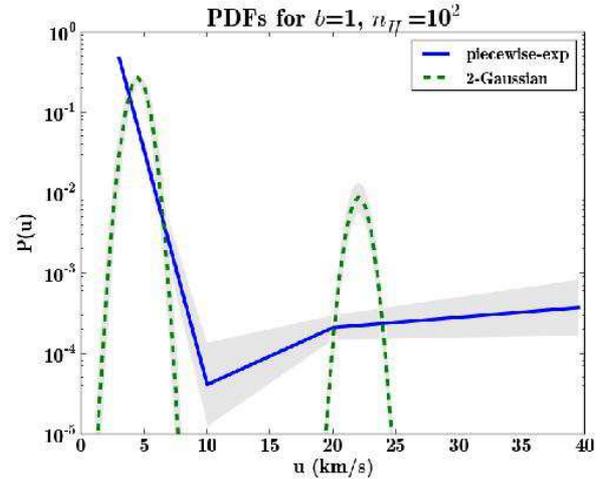}
\caption{Shape of our best two optimal PDFs fit to the SQ data. Grey areas represent uncertainties
due to the propagation of observational errors (see details in text).
\label{Flo:fit_fig4}}

\end{figure}

All optimal PDF shapes we find (including those with a large
$\chi^{2}$) are statistically biased towards low velocity shocks.
Figure \ref{Flo:fit_fig4} displays the two best fit PDF solutions. The
error bars (gray regions in these figures) are determined as follows:
we vary the parameters of the PDFs at the extreme of their 3-$\sigma$
range of uncertainty and we take the minimum and maximum values
predicted when using these extreme PDFs.  The piece-wise exponential
adjustment shows a dip at 10 km.s$^{-1}$ which indicates a bimodal
distribution of shocks consistent with the two-Gaussian fit (which is
bimodal in essence...).  Perturbation tests of this
 fit show that the PDF level at 10 km.s$^{-1}$ depends
mainly on the ratio S(0)/S(1), with a deeper dip for higher values of
this ratio. 
 On the other hand, the currently observed data do not
discriminate whether the distribution of moderate velocity shocks is wide
spread or well centred on a distinct velocity. Indeed, as seen on
figure \ref{Flo:b1.atomic_lines}(b), above 20 km.s$^{-1}$ the emission
properties of the observed H$_{2}$ lines do not change with velocity,
hence these lines cannot probe the shape of the PDF in this range of
velocities.  Ro-vibrational lines should be used to probe the shocks
at these velocities.
 We note all piecewise-exponential solutions
(including those with a larger $\chi^{2}$ at higher densities) are
similarly consistent with bimodal distributions. This is not 
expected from the statistics of shock velocities in numerical
simulations of turbulence.  It may be a result of the intrinsic
multiphase nature of the interstellar medium, where the moderate velocity
shocks would be interpreted as collisions between molecular clouds and
low velocity shocks would be the signature of turbulence dissipation
within these clouds.

For each of the optimal PDF found, we can predict the value $\hat{X}$
of other quantities of interest. We provide in table \ref{Flo:results}
 the total column densities for some molecules  
and the expected emission of some atomic lines in our best two solutions. {\bf These total column-densities are integrated from the pre-shock to the point where the temperature decreases back to 20\% above the pre-shock temperature.}   
In particular, we provide  a measure of the total H$_{2}$
column-density: the piece-wise exponential model yields 
$$N_{\mbox{shocked}}({\rm H}_{2})=2.6\,\times 10^{20}\mbox{cm}^{-2}$$
and the 2-Gaussian model yields 
$$N_{\mbox{shocked}}({\rm H}_{2})=1.0\,\times 10^{20}\mbox{cm}^{-2}\mbox{.}$$
Recent measurements from CO spectroscopy \citep{2012ApJ...749..158G} allow to estimate
the total H$_{2}$ column-density in the SQ large scale shock structure as 
$$N_{\mbox{total}}({\rm H}_{2})=8.5\,\times 10^{20}\,\pm1.5\,\times
10^{20}\,\mbox{cm}^{-2}$$ (assuming a Galactic value for the CO
emission to H$_2$ column conversion factor).  This puts the fraction
of shocked gas in this line of sight between 32\% and 12\%, depending
on which model we adopt, which confirms the results of
\citet{2010A&A...518A..59G} that quite substantial amounts of gas are
shock-heated. In fact, quite a few of our optimal solutions predict
$N_{\mbox{shocked}}({\rm H}_{2})>10^{21}\mbox{cm}^{-2}$ and are
therefore probably ruled out (only the viable solutions are underlined
in table \ref{Flo:chi2}).  

Since the shocks account for a significant fraction of the gas mass,
any chemical enhancements in the shocks may have a significant
contribution to molecular abundances. Table \ref{Flo:results} also
gives estimates of the chemical yields in these shocks. For example,
the H$_2$O abundance is on the order of 10$^{-6}$, significantly greater to that
observed in the diffuse ISM in the Milky-Way \citep{2010A&A...521L..34W}. These abundances
could be even higher if the medium is clumped (and therefore shielded) and {\it Herschel} may provide a
good opportunity to test this. For standard irradiation C$^+$ is the
dominant carbon gas species and this is also a result which will depend
on shielding.

The importance of these species (in addition to H$_2$),
as cooling agents of the interstellar turbulence depends directly on
their abundance. {\it Herschel} observations could thus provide additional
constraints to achieve a precise modelling of the physical and
chemical state of the gas.  We provide estimates of the total
integrated flux in the molecular cooling agents in table
\ref{Flo:results}: H$_2$ itself is responsible for more than 94\% of
the cooling in both our best-fit models. However its emission is distributed over many different
lines, the strongest of which H$_2$-S(1) contributes by a few percent
of the total. Note that for low-velocity C-shocks the pure rotational lines
sum up to almost all the H$_2$ cooling. On the contrary, moderate velocity J-shocks 
experience much higher temperatures and the energy in these shocks is radiated
away through higher H$_2$-levels. If our models with a contribution from moderate velocity J-shocks
hold, a lot more energy should be emitted
from higher excitation levels of the H$_2$ molecule than is observed from
pure rotational lines. 
Finally, note that other molecular coolants contribute at most a few percent
of the H$_2$-S(1) luminosity.

The emission of the C$^{+}$ line at 158 $\mu$m is
quite strong in our best two models (about half as strong as the
H$_{2}$-S(1) line), and so is the OI emission to a lesser extent.
Assuming all carbon is in the form of C$^+$ in the un-shocked fraction
of the total line of sight column-density $N_{\mbox{total}}({\rm
  H}_{2})$, we predict that the C$^{+}$ line should shine about as
much as the H$_{2}$-S(1) line, with a significant contribution from
shocks (66\% and 52\% respectively in both models). {\bf Interestingly, this was
also suggested in an interpretation of AKARI observations by \citet{2011ApJ...731L..12S}}. In contrast, the
model proposed by \citet{2009A&A...502..515G} with no UV irradiation
has no CII emission. This means that C$^{+}$ can indeed be a good
signature for the dissipation of kinetic and magnetic energy in
weakly shielded gas \citep[as previously found by
][]{2007IAUS..237...24F}. We also note that our viable models at
higher densities predict even stronger emission in C$^+$ by up to a
factor of three.  As such, C$^+$ emission measurements will help to
probe both the density and shielding of the gas. 
{\bf By contrast to C$^+$, the measured emission of the Si$^+$ line at $34.8~\mu$m is much greater than the predicted emission from our shock models. It is probably dominated by the contribution from the hot
  ionised medium, as already suggested by
  \citet{2010ApJ...710..248C}.}

\begin{figure}
\includegraphics[width=8.8cm]{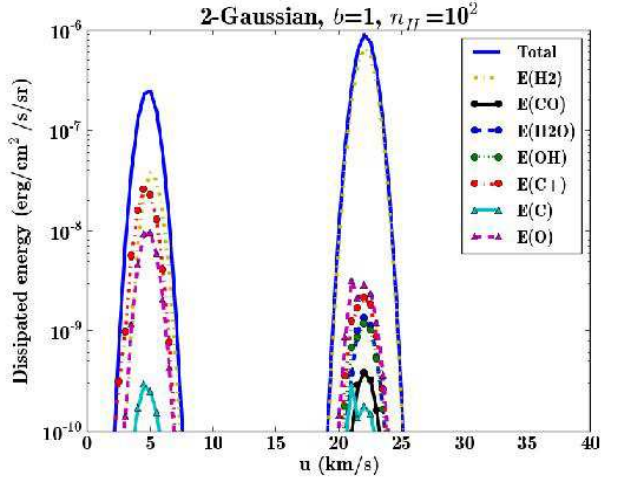}
\caption{Distribution of the energy dissipated via several cooling agents for the 2-Gaussian best-fit model.
\label{Flo:dissipated-energy}}

\end{figure}

Figure \ref{Flo:dissipated-energy} displays the distribution of energy radiated away
for several cooling agents in the best-fit 2-Gaussian
solution. A significant fraction of energy is dissipated in
low-velocity shocks but more energy is radiated within moderate velocity
shocks. Indeed the dissipated power goes as the cube of the velocity,
which slightly more than compensates the higher number of low-velocity
shocks. The figure clearly demonstrates that H$_2$ is by far the main
cooling agent. It also shows that the emission from atomic cooling
agents comes mainly from low-velocity shocks whereas the molecular
emission is dominated by moderate velocity shocks. Similarly, we have
checked that the emission from low $J$ H$_{2}$-lines probe
low-velocity shocks whereas higher $J$ lines probe larger velocity
shocks, which seems rather natural.

\begin{table}
\begin{tabular}{ccccccc}
$b$ &  $n_{\rm H}$& 1-Gauss & pow-law & exp. & pw-exp. & 2-Gauss\tabularnewline
\hline
\hline
0.1 & 10$^2$ & \underline{371.8 } & 2307.0  & 54.3  & \underline{60.8 } & 11.2    \tabularnewline
0.1 & 10$^3$ & \underline{504.0 } & 1650.4  & 152.4  & \underline{61.1 } & 105.6    \tabularnewline
0.1 & 10$^4$ & \underline{416.1 } & \underline{2139.9 } & 174.3  & \underline{580.8 } & 155.3    \tabularnewline 
   1 & 10$^2$ & 1628.5  & \underline{184.2 } & 598.5  & $\underline{\mathbf{2.6 }}$ & $\underline{\mathbf{2.0 }}$   \tabularnewline
   1 & 10$^3$ & 139.3  & 175.1  & 35.9  & \underline{5.0 } & \underline{13.8 }   \tabularnewline
   1 & 10$^4$ & 130.3  & 1648.0  & 12.6  & 6.3  & \underline{15.8 }  \tabularnewline
\end{tabular}

\caption{Summary of the optimal $\chi^{2}$ values obtained in all our attempts to fit the SQ data. Both
our optimal values are emphasised in bold faces. $n_{\rm H}$ is given in cm$^{-3}$. 
Underlined values correspond to solutions which have $N($H$_2)\leq 10^{21}$cm$^{-2}$.}
\label{Flo:chi2}
\end{table}

\begin{table}
\begin{tabular}{ccccc}
name  & pw-exp & 2-Gauss & obs. & err.\tabularnewline
\hline
\hline
\multicolumn{5}{c}{H$_2$-lines observations (erg.cm$^{-2}$.s$^{-1}$.sr$^{-1}$) }  \tabularnewline
\hline
0-0S(0) & 5.7(-7) & 5.6(-7) & 5.8(-7) & 3.5(-8)\tabularnewline
0-0S(1) & 4.2(-6) & 4.2(-6) & 4.3(-6) & 4.8(-8)\tabularnewline
0-0S(2) & 1.7(-6) & 1.7(-6) & 1.7(-6) & 7.0(-8)\tabularnewline
0-0S(3) & 3.8(-6) & 4.0(-6) & 4.2(-6) & 1.5(-7)\tabularnewline
0-0S(4) & 9.0(-7) & 9.4(-7) & 4.7(-7) & 1.8(-7)\tabularnewline
0-0S(5) & 2.5(-6) & 2.4(-6) & 2.6(-6) & 1.3(-7)\tabularnewline
\hline
\multicolumn{5}{c}{ Predictions} \tabularnewline
\hline
\multicolumn{5}{c}{Column-densities (cm$^{-2}$)}  \tabularnewline
\hline
H$_2$(0,0) & 5.8(19) & 2.2(19) & - & -   \tabularnewline
H$_2$(0,1) & 1.8(20) & 7.6(19) & - & -   \tabularnewline
H$_2$(0,2) & 3.4(18) & 3.4(18) & - & -   \tabularnewline
N(H$_2$) & 2.5(20) & 1.0(20) & <1.0(21) & -   \tabularnewline
N(H) & 4.3(19) & 1.8(19) & - & -   \tabularnewline
N(C+) & 7.6(16) & 3.2(16) & - & -   \tabularnewline
N(CO) & 2.5(14) & 1.8(14) & - & -   \tabularnewline
N(H$_2$O) & 1.6(14) & 1.9(14) & - & -   \tabularnewline
N(OH) & 1.9(14) & 1.7(14) & - & -   \tabularnewline
N(CH) & 2.1(13) & 1.0(13) & - & -   \tabularnewline
N(CH+) & 1.7(12) & 1.1(12) & - & -   \tabularnewline
N(HCO+) & 7.4(11) & 6.2(11) & - & -   \tabularnewline
\hline
\multicolumn{5}{c}{Atomic lines (erg.cm$^{-2}$.s$^{-1}$.sr$^{-1}$)} \tabularnewline
\hline
S(25.2m) & 1.4(-8) & 7.0(-9) & - & -   \tabularnewline
Si+(34.8m) & 9.8(-8) & 7.9(-8) & 1.6(-6) & 5(-8)   \tabularnewline
O(63.2m) & 1.4(-6) & 9.5(-7) & - & -   \tabularnewline
O(145.3m) & 1.2(-7) & 9.0(-8) & - & -   \tabularnewline
C+(158m) & 3.4(-6) & 2.2(-6) & - & -   \tabularnewline
C(370.4m) & 4.6(-8) & 3.1(-8) & - & -   \tabularnewline
C(609.8m) & 2.2(-8) & 1.3(-8) & - & -   \tabularnewline
\hline
\multicolumn{5}{c}{Integrated molecular cooling (erg.cm$^{-2}$.s$^{-1}$.sr$^{-1}$)}  \tabularnewline
\hline
E(H$_2$) & 1.6(-4) & 5.4(-5) & - & -  \tabularnewline 
E(CO) & 4.4(-8) & 2.9(-8) & - & -  \tabularnewline
E(H$_2$O) & 1.1(-7) & 1.0(-7) & - & -   \tabularnewline
E(OH) & 4.1(-7) & 1.0(-7) & - & -   \tabularnewline
\hline
\end{tabular}

\caption{Results of our best two fits to the observed spectral line energy distribution in the SQ (flux are given in erg.cm$^{-2}$.s$^{-1}$.sr$^{-1}$) and their respective predictions: column-densities are in cm$^{-2}$ and cooling is indicated in erg.cm$^{-2}$.s$^{-1}$.sr$^{-1}$. {\bf Parentheses denote powers of ten.}}
\label{Flo:results}
\end{table}

\subsubsection{Chamaeleon }

\citet{2008A&A...483..485N} successfully interpreted observational
results of a line of sight {\bf sampling diffuse molecular gas in front of the star HD 102065 in Chamaeleon.} Their PDR
model with $G_{0}=0.4$, $n_{\rm H}=80$
cm$^{-3}$ and $A_{v}=0.67$ reproduced accurately as many as seven independent observational
quantities.  However, they were unable to account for the
column-densities of rotationaly excited states of H$_{2}$ and failed
to reproduce the observed abundance of CH$^{+}$. Here, we assume that
both a PDR and a statistical distribution of shocks contribute to this line
of sight. Hence we attempt to fit as many as 11 observational measurements
available for this line of sight with a linear combination of our
shock models and a PDR model. 

 The PDR model is computed with our shock code in a PDR mode, see
 section \ref{Irradiation}. In principle, we should fit all
 irradiation parameters of the PDR alongside the grid of shocks. But
 this would assume we recompute the whole grid of shocks each time we
 probe new irradiation parameters and would lead to prohibitive
 computational time. Hence we fixed the irradiation conditions and the
 PDR model therefore introduces only one extra parameter in the fit:
 its weight in the line of sight compared to the shock models. We use
 $G_{0}=4$ and $A_v=0.335$ as entrance conditions for both the PDR and
 the shocks and we compute a new grid
 of shock models from 3 to 40 km.s$^{-1}$ with $n_{\rm H}=80$
 cm$^{-3}$ and an ortho/para ratio of 0.7 as is observed.

We thus obtain reasonable $\chi^{2}$ values with a best value of 11
for both the 2-Gaussian and the piecewise-exponential. 
 The error mainly comes from overestimating the column-density
in $J=5$, but the molecular chemistry of the line of sight is
reproduced rather accurately (see table \ref{tab:results-cham}).

The best two fitting PDFs compare qualitatively well to the solutions
we found for SQ, except the low velocity part of the 2-Gaussian is
more tightly peaked on low velocities (the moderate velocity component
accounts for 3\% of the total probability), and the
piecewise-exponential dip at 10 km.s$^{-1}$ is less pronounced. The two PDF
shapes are hence slightly less consistent than for the SQ, and
continuous solutions are not ruled out (the exponential and power-law
solutions yield $\chi^2$ values on the order of 20 and 15
respectively). However, in the case of the Chamaeleon, the physical
extent of the measured beam is much smaller than for the SQ. In
particular, the total number of shocks involved is presumably much
smaller and discrete numbers effects might show up in the
distribution.

 Our best two models predict similar values for most of the
 observables listed in table \ref{tab:results-cham}.  We now have two
 contributions from the PDR and the shocks and for each quantity
 listed in table \ref{tab:results-cham} we quote in brackets which
 fraction comes from the shocks. The weight of the PDR model in both
 best solutions is slightly different, and this yields a mass-fraction
 of shocked matter on the line of sight of 4\% in the
 piecewise-exponential model and 19\% in the 2-Gaussians model. The
 contribution from shocks is hence small but not
 insignificant. However, molecules other than H$_2$ are all formed in
 the shocks and not in the PDR, with for example 98\% of the CO coming
 from the shocks.  Note that in both our two best models, shocks with
 velocity above 15 km.s$^{-1}$ account for the formation of almost all
 molecules. This is due to the overcome of chemical barriers thanks to
 the thermal energy released in shocks and to the ion-neutrals drift,
 as mentionned in sections \ref{modelsJ} and \ref{modelsC} .

 CH$^+$ needs mechanical energy injection to raise {\bf the effective} temperature for its
 formation which is why pure PDR models failed to account for its
 abundance.  In fact, CH$^+$ forces a too hot component into the
 model, which then over-produces the $J=5$ H$_2$-level.  A possible
 solution would be that the irradiation in the far-UV is enhanced in
 the Chamaeleon compared to the Draine ISRF: this would make more
 CH$^+$ from CH$_2^+$ and CH$_3^+$ photo-dissociation \citep{2010A&A...518L.118F} while requiring
 a lower temperature more compatible with the observed $J=5$
 H$_2$-level column-density.  In fact, UV spectra in
 \citet{1994A&A...284..956B} suggest that the enhanced far-UV field
 could be accounted by a flatter extinction curve in the far-UV for
 this very line of sight.

 As for the SQ, H$_2$ is by far the main cooling agent (it radiates more than 92\%
of the total cooling in both best-fit models) but
 CII (and OI to a lesser extent) has emission comparable to the
 H$_2$-S(1) line. Even though it is now dominated by the emission from
 the PDR, the shock contribution to CII emission is still significant (from
 18\% to 31\% according to both best models). Interestingly, we
 predict OH should shine about as much as CII and as for the SQ solution, the OH emission is
 almost completely dominated by the contribution from shocks at moderate
 velocity. OH observations would hence provide a good test of the
 existence or not of a moderate velocity shock component in the
 Chamaeleon. 

\begin{table}
\begin{tabular}{ccccc}
names & piecewise-exp   & 2-Gaussian  & obs & error   \tabularnewline 
\hline 
\hline 
\multicolumn{3}{c}{Column-densities (cm$^{-2}$)} &   \tabularnewline
\hline
 & \multicolumn{2}{c}{Observations} &   \tabularnewline
\hline
N(H) & 2.6(20) $[0.06]$ & 2.7(20) $[0.23]$ & 3.1(20) & 3.0(19) \tabularnewline 
H$_2$(0,0) & 1.8(20) $[0.04]$ & 1.8(20) $[0.20]$ & 2.0(20) & 2.0(19) \tabularnewline 
H$_2$(0,1) & 2.1(20) $[0.03]$ & 1.9(20) $[0.14]$ & 1.4(20) & 1.0(19) \tabularnewline 
H$_2$(0,2) & 2.6(18) $[0.53]$ & 2.6(18) $[0.61]$ & 2.6(18) & 1.0(17) \tabularnewline 
H$_2$(0,3) & 4.2(17) $[0.99]$ & 4.9(17) $[0.99]$ & 2.0(17) & 1.0(17) \tabularnewline 
H$_2$(0,4) & 4.4(16) $[1.00]$ & 5.9(16) $[1.00]$ & 2.0(16) & 1.0(16) \tabularnewline 
H$_2$(0,5) & 2.2(16) $[1.00]$ & 3.8(16) $[1.00]$ & 7.1(14) & 5.0(14) \tabularnewline 
N(C) & 2.4(15) $[0.20]$ & 2.3(15) $[0.30]$ & 6.0(14) & 1.5(14) \tabularnewline 
N(CO) & 5.2(13) $[0.98]$ & 4.9(13) $[0.98]$ & 5.6(13) & 6.8(12) \tabularnewline 
N(CH) & 6.2(12) $[0.87]$ & 4.8(12) $[0.86]$ & 6.4(12) & 1.0(12) \tabularnewline 
N(CH+) & 8.8(12) $[1.00]$ & 1.0(13) $[1.00]$ & 1.2(13) & 1.9(12) \tabularnewline 
\hline 
 & \multicolumn{2}{c}{Predictions} &   \tabularnewline
\hline
N(H$_2$) & 3.9(20) $[0.04]$ & 3.8(20) $[0.17]$ & - & - \tabularnewline 
N(C+) & 1.4(17) $[0.05]$ & 1.4(17) $[0.19]$ & - & - \tabularnewline 
N(H$_2$O) & 2.8(13) $[0.99]$ & 5.3(13) $[1.00]$ & - & - \tabularnewline 
N(OH) & 6.9(13) $[0.99]$ & 1.1(14) $[0.99]$ & - & - \tabularnewline 
N(HCO+) & 2.7(11) $[0.99]$ & 3.4(11) $[0.99]$ & - & - \tabularnewline 
N(CN) & 3.9(11) $[0.99]$ & 5.3(11) $[0.99]$ & <5.9(11) & - \tabularnewline 
N(C$_2$) & 3.6(10) $[0.80]$ & 2.8(10) $[0.77]$ & <1.4(13) & - \tabularnewline 
\hline 
\multicolumn{3}{c}{Atomic lines (erg.cm$^{-2}$.s$^{-1}$.sr$^{-2}$)} &   \tabularnewline
\hline
S(25.2m) & 4.5(-9) $[1.00]$ & 5.1(-9) $[1.00]$ & - & - \tabularnewline 
Si+(34.8m) & 6.4(-8) $[0.82]$ & 6.0(-8) $[0.84]$ & - & - \tabularnewline 
O(63.2m) & 1.2(-6) $[0.66]$ & 1.1(-6) $[0.67]$ & - & - \tabularnewline 
O(145.3m) & 8.7(-8) $[0.76]$ & 7.7(-8) $[0.77]$ & - & - \tabularnewline 
C+(158m) & 3.8(-6) $[0.18]$ & 3.9(-6) $[0.31]$ & 2.8(-6) & - \tabularnewline 
C(370.4m) & 2.2(-8) $[0.60]$ & 2.3(-8) $[0.68]$ & - & - \tabularnewline 
C(609.8m) & 1.3(-8) $[0.33]$ & 1.3(-8) $[0.44]$ & - & - \tabularnewline 
\hline
\multicolumn{3}{c}{Molecular cooling (erg.cm$^{-2}$.s$^{-1}$.sr$^{-1}$)} &  & \tabularnewline
\hline
0-0S(1) & 1.9(-6) $[0.99]$ & 2.2(-6) $[0.99]$ & - & -   \tabularnewline 
E(H$_2$) & 9.5(-5) $[1.00]$ & 5.7(-5) $[0.99]$ & - & - \tabularnewline 
E(CO) & 9.3(-8) $[1.00]$ & 8.5(-8) $[1.00]$ & - & - \tabularnewline 
E(H$_2$O) & 1.8(-7) $[1.00]$ & 2.3(-7) $[1.00]$ & - & - \tabularnewline 
E(OH) & 2.9(-6) $[1.00]$ & 2.0(-6) $[1.00]$ & - & - \tabularnewline 
\hline
\end{tabular}

\caption{Results and predictions of the fit of  piecewise-exponential to the Chamaeleon
data \citep{2002A&A...391..675G,2008A&A...483..485N}. {\bf Powers of ten are indicated in parentheses.} For each value, we provide the
fraction coming from the shocks in brackets.\label{tab:results-cham}}

\end{table}

\section{Summary, conclusions, discussion, prospects}

We provide to the scientific community a grid of shock models 
\footnote{The output data of our models is archived on
  http://cemag.ens.fr } in UV heated gas which may be used to
interpret observations of the Milky-Way diffuse ISM and integrated
properties of galaxies in general.  We examine magnetised shocks in
media with densities from $n_{\rm H}=10^{2}\mbox{ to
}10^{4}\,\mbox{cm}^{-3}$, with standard ISM irradiation conditions and
at low to moderate velocities (from 3 to 40 km.s$^{-1}$). When the
velocity of these shocks is below the critical velocity for the
existence of pure C-type shocks, we compute C-type shocks, otherwise,
we compute J-type shocks. The models include the effects of the
ambient UV irradiation field on the pre-shock chemistry and thereby on
the relative importance of cooling lines in the shock.  For instance,
C$^{+}$ can provide significant line cooling in shocks propagating in
the UV irradiated gas where it is the dominant carbon species.

We illustrate how the model results may be used to interpret data
on one galactic line of sight through the diffuse ISM (Chamaeleon)
and one extra-galactic object (the SQ shock). We fit the observables
with a continuous combination of shock models as a phenomenological
description of the complex statistical properties of the dissipation
of turbulence. The model provides a rather good match to the data
 for bimodal
distributions of shock velocities. This does not match predictions from numerical simulations
of \citet{2000A&A...362..333S} who found a power-law distribution with an exponential cut-off.
 Further work is needed to understand the
reasons behind this apparent discrepancy. Interestingly, the low and
moderate velocity components of our best-fit PDF operate in very different
regimes of energy dissipation. Indeed, in our best-fit models to the observations, the low velocity shocks are
of C-type and they dissipate energy via ion-neutral drift whereas
the larger velocity shocks are of J-type and they undergo viscous
dissipation. 

In both our interpretations of the SQ and the Chamaeleon, a
significant fraction of the molecular gas is shock heated and the
chemistry in this shock heated gas has a dominant contribution to key
molecules such as CO, H$_{2}$O, OH or CH$^{+}$ which are commonly used
as diagnostics. This shock heated contribution might be greater than
the contribution of CO computed in simulations by
\citet{2011MNRAS.412..337G} where the resolution in the shocks is too
scarce to trigger the shock chemistry:  {\bf  too low resolution
 smears out the temperature to such an extent that shocks become
 nearly isothermal with a temperature too low for molecule
 formation.} 
Furthermore, although low-velocity shocks
are less efficient than moderate velocity shocks, they are more
numerous and depending on the statistical distribution of shocks they 
could potentially account for a higher fraction of the excitation and
formation of molecules.

  The presence of shocks in the line of sight also has important consequences on the emission
of C$^{+}$. Indeed, C$^{+}$ is usually considered as a good tracer
of the ambient UV radiation field, and indeed the mere presence of
the C$^{+}$ ion is conditioned to its photo-ionisation. But we show
here that the emission of C$^{+}$ in both the SQ and Chamaeleon 
has a significant contribution from 
gas heated by the dissipation of kinetic energy rather than by UV
photons as is the case in photon-dominated regions (PDR). 

The SQ H$_{2}$ excitation diagram was fit with low and moderate velocity C-shocks in
dense UV-shielded gas ($n_{\rm H}=10^4$~cm$^{-3}$, $G_0=0$) in earlier
studies \citep{2009A&A...502..515G}. We show that a better fit can be
obtained with diffuse irradiated gas ($n_{\rm H}=10^2$~cm$^{-3}$,
$G_0=1$) including J-shocks. If this second solution is the appropriate one, C$^{+}$ is a
significant coolant of the shocked medium. This prediction can be
tested with the {\it Herschel} observatory. The presence of the high-velocity 
J-shocks can also be tested through the emission in H$_2$-lines coming
from rovibrational levels.

OI, H$_{2}$O, OH and CO provide additional diagnostic lines which
could help constraining the physical state of the gas and the dissipation
processes. OI is seen to be significant at larger densities whereas
C$^{+}$ is significant for lower densities. Molecules such as H$_{2}$O,
OH and CO evacuate less than a percent of the total kinetic power
except at large densities where H$_{2}$O in C-shocks
can contribute to as much as 10\%. 

The remainder of the incoming power in our shocks is converted into
a flux of magnetic energy. The differences in total pressure (magnetic plus
thermal) between the shocked regions and the regions which have not
been shocked will initiate new motions: the energy stored in the post-shock
compression will eventually give rise to expansion motions, with an
effective transfer back to kinetic energy. Thus, shocks are actors
which contribute to the equipartition between various forms of energy,
thanks to a continuous recycling of the gas throughout successive
shock structures.

The main caveat of our study is that we assume the relaxation
of the post-shock pressure takes place instantaneously: namely, our
model accounts for the pre-shock gas and the shocked gas but does
not incorporate the expansion of the gas after the shock. We shall
attempt to model this phase in future studies. {\bf Another caveat of our study lies in the 
implicit assumption that all shocks are steady. Indeed some shock velocities 
are prone to instabilities and even quasi-steady models are invalid in this case 
(see \citealt{2004A&A...427..157L}, for example).} Less important caveats are the shock
orientation, its curvature, the neglect of intermediate ages of shocks
(for ages lower than about 10$^{4}$ yr, we should consider CJ-type
shocks or non-steady shocks). We believe that these minor caveats
are washed out by the fact that we fit a statistical collection of
shock velocities. Finally we do not treat UV pumping in the H$_{2}$-population
but \citet{1988MNRAS.234..863M} showed this starts to make a difference
only for H$_{2}$ levels above $J=5$ and for $G_{0}\gg1$. 

Our PDF fitting technique might prove useful to account for single
shocks with more complicated geometries. For example, a single bow-shock
effectively encompasses a continuous collection of velocities depending
on the angle at which each fluid parcel of pre-shock gas impinges
on the bow-shock. In such a framework, the shape of the lines is directly
related to the dynamics inside the shock, and not simply due to the
relative motions between each shocks as in the statistical distribution
of shocks we infer in the SQ. In such cases, it will therefore be
interesting to predict and use the line shapes as additional observational
constraints on the shock physics.

Finally, as we already mentioned, the dissipation of turbulent motions does not
occur only in shocks. It will be interesting in future work to assess
on the one hand what fraction of the energy dissipation takes place
in vortices, current sheets or shocks, for example. On the other hand,
it will be worth comparing vortex models as in \citet{2009A&A...495..847G}
to our own shock models in order to check whether we can disentangle
observationally the respective signatures of shocks and vortices.

\begin{acknowledgements}
  P.L., E.F. and B.G. acknowledge support from SCHISM A.N.R.. We thank Maryonne
  Gerin and Antoine Gusdorf for comments on the manuscript. We thank the anonymous referee for his comments which improved the readability of the paper.
\end{acknowledgements}

\bibliographystyle{aa}
\addcontentsline{toc}{section}{\refname}\bibliography{biblio}

\end{document}